\documentclass[aps,onecolumn,superscriptaddress, nofootinbib, longbibliography, notitlepage]{revtex4-1}  

\usepackage{amssymb,amsmath,amsfonts,amsbsy}
\usepackage{color}
\usepackage{enumerate}
\usepackage{graphicx}
\usepackage{pifont}

\usepackage{subfigure}
\usepackage{multirow}
\usepackage{cancel}

\RequirePackage[colorlinks,citecolor=blue,urlcolor=blue,linkcolor=blue]{hyperref}

\usepackage{xcolor}
\usepackage[normalem]{ulem} 

\usepackage{float}
\usepackage[export]{adjustbox}

\newcommand{\be}{\begin{equation}}
\newcommand{\ee}{\end{equation}}
\newcommand{\bea}{\begin{eqnarray}}
\newcommand{\eea}{\end{eqnarray}}

\begin{document}

\title{Stochastic gravitational wave background phenomenology in a pulsar timing array}

\author{Reginald Christian Bernardo}
\email{rbernardo@gate.sinica.edu.tw}
\affiliation{Institute of Physics, Academia Sinica, Taipei 11529, Taiwan}

\author{Kin-Wang Ng}
\email{nkw@phys.sinica.edu.tw}
\affiliation{Institute of Physics, Academia Sinica, Taipei 11529, Taiwan}
\affiliation{Institute of Astronomy and Astrophysics, Academia Sinica, Taipei 11529, Taiwan}

\begin{abstract}
Pulsar timing offers an independent avenue to test general relativity and alternative gravity theories. This requires an understanding of how metric polarizations beyond the familiar transverse tensor ones imprint as a stochastic gravitational wave background and correlate the arrival time of radio pulses from a pair of millisecond pulsars. In this work, we focus on an isotropic stochastic gravitational wave background and present a straightforward, self-contained formalism for obtaining the power spectrum and the overlap reduction function, the relevant physical observable in a pulsar timing array, for generic gravitational degrees of freedom featuring both transverse and longitudinal modes off the light cone. We additionally highlight our consideration of finite pulsar distances, which we find significant in two ways: first, making all the modes well defined, and second, keeping the small scale power that is contained by pulsars of subdegree separations in the sky. We discuss this for tensor, vector, and scalar polarizations, for each one focusing on the angular power spectrum and the overlap reduction function for an isotropic stochastic gravitational wave background. Our results pave the road for an efficient numerical method for examining the gravitational wave induced spatial correlations across millisecond pulsars in a pulsar timing array.
\end{abstract}

\maketitle


\section{Introduction}
\label{sec:intro}

The groundbreaking direct discovery of gravitational waves by the LIGO/Virgo collaborations was probably the last decade's most significant scientific breakthrough \cite{LIGOScientific:2016aoc}. This has ushered in the era of gravitational wave astronomy, opening up a novel observational window into the Universe and letting in new independent opportunities to test our fundamental understanding of nature. The first few years alone have already made an outstanding impact, telling us that the graviton cannot be heavier than $10^{-22}$ electron volts and that gravitational waves -- spacetime distortions that displace masses on its path -- should practically be as fast as light in vacuum \cite{LIGOScientific:2017vwq, LIGOScientific:2021djp, LIGOScientific:2021sio}. These observations in the hundred to kilo hertz gravitational wave band will soon be supported by space based observations, relieved of the terrestrial restrictions, that aim to probe the millihertz frequencies \cite{LISA:2022kgy}. The science that gravitational wave observations promises, from learning their sources to the physics behind them, is simply astonishing and captures the interest and imagination of both the scientific community and the public alike.

At the same time, pulsar timing brings in another piece into the picture, providing gravitational wave observations in the nanohertz frequency band. This is done by observing the arrival time of radio pulses by millisecond pulsars which should be spatially correlated due to the stochastic gravitational wave background \cite{Burke-Spolaor:2018bvk}. The targeted sources here are phase transitions in the early universe, cosmic strings, and supermassive binary black holes, all of which carry information about the cosmological history \cite{Romano:2019yrj}. Recently, pulsar timing array efforts by the North American Nanohertz Observatory for Gravitational Waves (NANOGrav) \cite{NANOGrav:2020bcs}, the Parkes Pulsar Timing Array (PPTA) \cite{Shannon:2015ect}, the European Pulsar Timing Array (EPTA) \cite{Lentati:2015qwp}, or jointly the International Pulsar Timing Array (IPTA) which take in years of observation of about a hundred millisecond pulsars in the sky have presented strong evidence for a common spectrum process. This teases an intriguing departure from the quadrupolar dominated spatial correlation sourced by the transverse tensor modes expected in general relativity \cite{Hellings:1983fr, NANOGrav:2020bcs, Chen:2021wdo, Chen:2021ncc}. Granted, the uncertainties in the present data set are quite large due to the limitations in the optimal statistic analysis, but this also advances alternative scenarios where nontensorial spatial correlations weigh in the stochastic gravitational wave background. Perhaps tensor modes off the light cone, say, by a dispersive gravitational wave \cite{deRham:2012az}, or additional gravitational degrees of freedom such as the ones that come in modified gravity theories \cite{Chamberlin:2011ev, Liang:2021bct, Tachinami:2021jnf, Odintsov:2021kup, Oikonomou:2022xoq} can explain this correlation. Unequivocally, the science we have yet to learn about the nanohertz gravitational wave sky is rich and exquisite \cite{Tasinato:2022xyq, Chu:2021krj, Bernardo:2022vlj}.

In this work, we study an isotropic stochastic gravitational wave background, utilizing a Sachs-Wolfe line of sight integral, in the setting of a pulsar timing array. We do so by directly constructing the power spectrum through which the overlap reduction function (ORF), describing the cross correlated power across pulsars in the sky, the main physical observable of interest, can be obtained \cite{Dai:2012bc, Qin:2018yhy, Qin:2020hfy, Liu:2022skj, Ng:2021waj}. We argue this alternative to the so called real space formalism \cite{NANOGrav:2021ini, Boitier:2021rmb, Hu:2022ujx} is numerically efficient for the present data set, requiring only the first few multipoles since each pulsar pair has at least about a {three} degree separation. We further take into consideration finite pulsar distances and arbitrary propagation velocities, which make \textit{all} of the metric polarizations well defined unlike their infinite distance and luminal counterparts. To the best of our knowledge, this is the most general setup that take finite pulsar distances and arbitrary velocities in the literature. Our main results are presented in Section \ref{sec:summary} for future data analysis work, whereas the rest of the paper is spent on the derivation and a discussion of the phenomenological signatures of an isotropic stochastic gravitational wave background in a pulsar timing array.
 
After the summary (Section \ref{sec:summary}), we lay down the formalism we follow in calculating the stochastic gravitational wave background's power spectrum and overlap reduction function (Section \ref{sec:ptobservables}). This includes a brief review of the real space formalism which we also use to confirm our power spectrum calculations. Then, we derive the power spectrum and the overlap reduction function for tensor (Section \ref{sec:tensor_pols}), vector (Section \ref{sec:vector_pols}), and scalar (Section \ref{sec:scalar_pols_finite_dist}) metric polarizations, and for each one study their phenomenology. We discuss the advantages and disadvantages of the power spectrum calculation and how it can be improved (Section \ref{sec:discussion}). We remark on several future work including possible anisotropies in the stochastic gravitational wave background (Section \ref{sec:conclusions}). In the appendices, we explicitly write down the polarization tensors considered in our analysis (Appendix \ref{sec:gw_pol_tensors}) and list down identities for spherical harmonics relevant to the calculations in the main text (Appendix \ref{sec:3Y3j}). We also briefly review another approach to obtaining the overlap reduction function (Section \ref{sec:another}) and outline some of the well known analytical results in the literature (Section \ref{sec:orfs_review}).

We work with the mostly plus metric signature $(-, +, +, +)$ and geometrized units $c = 8 \pi G = 1$. Also, we list down the symbols often mentioned in this paper in Table \ref{tab:symbols}.

\begin{table}[h!]
    \centering
    \caption{Description of the common symbols appearing throughout this paper.}
    \begin{tabular}{|c|c|}
    \hline
    symbol & description  \\ \hline \hline
    $\gamma_{ab}(\zeta)$ & overlap reduction function \\ \hline
    $\zeta$ & angular separation \\ \hline
    $\gamma_{ab}(0)$ & overlap reduction function at $\zeta = 0^\circ$ (`zero lag') \\ \hline
    $\gamma_{aa}$ & autocorrelation function \\ \hline
    $l$ & multipole number/index \\ \hline
    $C_l$ & power spectrum multipole \\ \hline
    $D_a$ & distance to pulsar $a$ \\ \hline
    $f$ & gravitational wave frequency \\ \hline
    $v$ & group velocity \\ \hline
    $P_l(x)$ & Legendre polynomial \\ \hline
    $j_l(x)$ & spherical Bessel function \\ \hline
    $Y_{lm}\left(\hat{k}\right)$ & spherical harmonics \\ \hline
    $\, _s Y_{lm}\left(\hat{k}\right)$ & spin weighted spherical harmonics \\ \hline
    $h_{ij}$ & gravitational wave \\ \hline
    $\vec{k} = k \hat{k}$ & wave vector \\ \hline
    $\varepsilon_{ij}$ & gravitational wave polarization basis tensor \\ \hline
    $\hat{e}_a$ & unit vector from earth to pulsar $a$ \\ \hline
    $d^{ij}$ & detector tensor $\hat{e}^i \otimes \hat{e}^j$ \\ \hline
    $z(t, \hat{e})$ & redshift fluctuation \\ \hline
    $r(t, \hat{e})$ & pulsar timing residual \\ \hline
    $F_a^A \left(\hat{k}\right)$ & antenna pattern for polarization $A$ \\ \hline
    $D_{m'm}^l(-\alpha,-\theta,-\phi)$ & Wigner D matrix \\ \hline
    $\Gamma(x)$ & gamma function \\ \hline
    $\, _2F_1(a, b;c;x)$ & hypergeometric function \\ \hline
    $\, _2\tilde{F}_1(a,b;c;x)$ & regularized hypergeometric function $\, _2F_1(a, b;c;x)/\Gamma(c)$ \\ \hline
    $\, _2F_2(a, b;c, d;x)$ & hypergeometric function \\ \hline
    $\, _2\tilde{F}_2(a,b;c, d;x)$ & regularized hypergeometric function $\, _2F_1(a, b;c, d;x)/\left( \Gamma(c) \Gamma(d) \right)$ \\ \hline
    \end{tabular}
    \label{tab:symbols}
\end{table}

\section{Summary of main results}
\label{sec:summary}

We summarize the main results of this paper intended for future work with spatial correlation data from a pulsar timing array.

The overlap reduction function, or rather the cross correlated power between pulsars in the sky, sourced by an isotropic stochastic gravitational wave background is given by
\begin{equation}
\label{eq:orf1}
    \gamma_{ab}\left( \zeta \right) = \sum_{l} \dfrac{2l + 1}{4\pi} C_l P_l \left( \cos \zeta \right) \,,
\end{equation}
where $\zeta$ is the angle between a pulsar pair and the $C_l$'s are the power spectrum multipoles. The gravitational degrees of freedom determine which polarizations contribute to $C_l$. We write this formally as
\begin{equation}
\label{eq:Clgen}
    C_l^A = \dfrac{J_l^A\left(f D_a\right) J_l^{A*}\left(f D_b\right)}{\sqrt{\pi}} \,,
\end{equation}
where the superscript index $A$ determines the mode contributing to the gravitational wave, $f$ is the gravitational wave frequency, and $D_i$ is the distance to the pulsars. This is an efficient formula in practice since only a small number of multipoles need to be considered. To elucidate on this point, consider for example, the smallest pulsar pair separation in NANOGRAV's 12.5 year data set which is about {three} degrees. In this case, even the first {sixty} multipoles, resolving angular separations $\zeta {\gtrsim} 180^\circ/l_\text{max} = {3}^\circ$, are sufficient for the analysis.

We enumerate in what follows the $J_l^A(x)$'s appearing in \eqref{eq:Clgen} for modes propagating at a velocity $v$. For the tensor polarizations, the $J_l^A(x)$'s are given by
\begin{equation}
\label{eq:JT}
    J^\text{T}_l\left(fD\right) = \sqrt{2} \pi i^l \sqrt{\dfrac{(l + 2)!}{(l - 2)!}} \int_0^{2\pi fDv} \dfrac{dx}{v} \ e^{ix/v} \dfrac{j_l(x)}{x^2} \,,
\end{equation}
while for the vector polarizations these are 
\begin{equation}
\label{eq:JV}
    J^\text{V}_l\left(fD\right) = 2 \sqrt{2} \pi i^l \sqrt{l(l+1)} \int_0^{2\pi fDv} \dfrac{dx}{v} \ e^{ix/v} \dfrac{d}{dx} \left( \dfrac{j_l(x)}{x} \right) \,,
\end{equation}
where $j_l(x)$ is the spherical Bessel function of the first kind. On the other hand, for the scalar transverse polarization, this is given by
\begin{equation}
\label{eq:JST}
    J^{\text{ST}}_l\left( fD \right) = 2 \sqrt{2} \pi i^l \int_0^{2\pi f D v} \dfrac{dx}{v} \ e^{ix/v} \left( j_l''(x) + j_l(x) \right) \,,
\end{equation}
and for the scalar longitudinal polarization, this is
\begin{equation}
\label{eq:JSL}
    J^{\text{SL}}_l\left( fD \right) = - 2 \pi i^l \int_0^{2\pi f D v} \dfrac{dx}{v} \ e^{ix/v} j_l(x) \,.
\end{equation}
The above integrals admit analytical expressions in the infinite distance limit (see for instance \cite{Qin:2020hfy}). In general, for finite pulsar distances, $f D_i < \infty$, and subluminal velocities $v < 1$, these integrals can be evaluated using computer algebra systems such as python and Mathematica, leading to the power spectrum, and the overlap reduction function.

A simple algorithm for spatial correlation data analysis is shown below:
\begin{enumerate}
    \item Select a gravitational wave mode $A$ and provide its velocity $v$ and frequency $f$;
    \item Calculate the power spectrum \eqref{eq:Clgen} given the set of pulsar distances $D_i$;
    \item Calculate the overlap reduction function $\gamma_{ab}^A(\zeta)$ using \eqref{eq:orf1};
    \item Obtain the normalized overlap reduction function $\Gamma_{ab}^A(\zeta) = 0.5 \times \gamma_{ab}^A(\zeta)/\gamma_{ab}^\text{HD}(0)$ where $\gamma^\text{HD}_{ab}(0) = \gamma_{ab}^\text{T}\left(\zeta = 0^\circ\right)|_{fD \rightarrow \infty, v \rightarrow 1}$;
    \item Compare the curve $A_\text{GW}^2 \Gamma_{ab}^A(\zeta)$ with the observed $\left[ A_\text{GW}^2 \Gamma_{ab}^A(\zeta) \right]_\text{PTA}$ from pulsar timing array, where $A_\text{GW}$ is the characteristic gravitational wave strain.
\end{enumerate}
These steps are easy to implement and, as we advocate for the current data set, requires only about {sixty} multipoles, making the computation very fast \cite{Bernardo:2022vlj}. We mention that the normalization (step 4) is an aesthetic choice. Overlap reduction functions used for data analysis are normalized relative to the traditional Hellings-Downs curve as $\Gamma^A_{ab}(0)$, such that $\Gamma_{ab}^\text{HD}(0) = 0.5$. {We put all these steps together with a calculation of the variances of the ORF \cite{Bernardo:2022xzl} in a python code `\href{http://ascl.net/2211.001}{PTAfast}' \cite{PTAfast}.}

We spend the bulk of this paper deriving the above equations in detail and studying their phenomenology in a pulsar timing array. We emphasize that this was not the first time the power spectrum approach was marketed for pulsar timing array analysis (see \cite{Dai:2012bc, Qin:2018yhy, Qin:2020hfy, Liu:2022skj}). Instead we present the most general output from this, keeping the pulsars at finite distances throughout and the propagation velocities arbitrary. Our derivation is in addition self contained and arguably pedagogical, requiring only a few textbook material on spherical harmonics and Bessel functions \cite{Weinberg, Arfken}. 

\section{Pulsar timing observables}
\label{sec:ptobservables}

We briefly discuss the pulsar timing residual and overlap reduction function \cite{Liu:2022skj, NANOGrav:2021ini}.

\subsection{Pulsar timing and gravitational waves}
\label{subsec:pulsar_timing_gws}

We consider a gravitational wave propagating along the $\hat{k}$ direction in a mixture of various polarizations $A$. In this context, for concreteness, `polarization' means the various independent ways a gravitational wave displaces masses on its path -- scalar transverse, scalar longitudinal, transverse vector, and transverse-traceless $(+, \times)$ tensor modes -- that is induced by the propagating degrees of freedom. In symbols, we represent this as a typical plane wave superposition,
\begin{equation}
\label{eq:gw_general}
    h_{ij}\left(\eta, \vec{x}\right) = \sum_A \int_{-\infty}^\infty df \int_{S^2} d\hat{k} \ h_A\left(f, \hat{k}\right) \varepsilon_{ij}^A e^{-2\pi i f \left( \eta - v \hat{k} \cdot \vec{x} \right)} \,,
\end{equation}
where $\varepsilon_{ij}^A$ are basis polarization tensors (Appendix \ref{sec:gw_pol_tensors}) and $v = d\omega/dk$ is the group velocity.

Now, the main observable in a pulsar timing experiment is the timing residual $r(t)$. Our goal is to single out the influence of a gravitational wave on this observable. We proceed to do this through the power spectrum.

To start, in terms of the redshift space fluctuation $z(t)$, we write down the timing residual as
\begin{equation}
\label{eq:timing_residual}
    r\left(t\right) = \int_0^t dt' \ z\left(t'\right) \,,
\end{equation}
where $t$ is the duration of an observation. For a passing gravitational wave $h_{ij}\left(\eta, \vec{x}\right)$, the redshift fluctuation considering a photon emitted at time $\eta_e$ and received by the detector at time $\eta_r$ is given by
\begin{equation}
\label{eq:z_swolf}
    z\left(t', \hat{e}\right) = - \dfrac{1}{2} \int_{t' + \eta_e}^{t' + \eta_r} d\eta d^{ij} \partial_\eta h_{ij} \left( \eta, \vec{x} \right) \,,
\end{equation}
where $d^{ij} = \hat{e}^i \otimes \hat{e}^j$ is the detector tensor with $\hat{e}$ being a unit vector pointing toward the pulsar from earth, in words, the projections along the pulsar's line of sight. Substituting the gravitational wave \eqref{eq:gw_general}, we have
\begin{equation}
    r\left(t, \hat{e}\right) = \int_0^t dt' \left( - \dfrac{1}{2} \right) \int_{t' + \eta_e}^{t' + \eta_r} d \eta \ d^{ij} \sum_{A} \int_{-\infty}^\infty df \int_{S^2} d\hat{k} \ h_A \left(f, \hat{k}\right) \varepsilon_{ij}^A\left(\hat{k}\right) \left(-2\pi i f\right) e^{-2\pi i f \left( \eta - v \hat{k} \cdot \vec{x} \right)} \,.
\end{equation}
We move forward by expanding the plane wave in terms of spherical harmonics, $Y_{lm}\left(\hat{e}\right)$,
\begin{equation}
    e^{2\pi i f v \hat{k}\cdot\vec{x}} = 4\pi \sum_{lm} i^l j_l\left(2\pi fv|\vec{x}|\right) Y_{lm}^*\left(\hat{k}\right) Y_{lm}\left(\hat{e}\right) \,,
\end{equation}
such that
\begin{equation}
\begin{split}
    r\left(t, \hat{e}\right) = & \int_0^t dt' \left( - \dfrac{1}{2} \right) \int_{t' + \eta_e}^{t' + \eta_r} d \eta \ d^{ij} \sum_{A} \int_{-\infty}^\infty df \int_{S^2} d\hat{k} \\
    & \ \ \times h_A \left(f, \hat{k}\right) \varepsilon_{ij}^A\left(\hat{k}\right) \left(-2\pi i f\right) e^{-2\pi i f \eta} \ 4\pi \sum_{lm} i^l j_l\left(2\pi fv \left( t' + \eta_r - \eta \right) \right) Y_{lm}^*\left(\hat{k}\right) Y_{lm}\left(\hat{e}\right) \,,
\end{split}
\end{equation}
where $\vec{x} = |\vec{x}| \hat{e} = \left( t' + \eta_r - \eta \right) \hat{e}$ is the position vector to the pulsar at time $t'$ and $j_l(x)$ is the spherical Bessel function of the first kind. We consider the following integral identities to simplify this:
\begin{equation}
    \int_0^t dt' \int_{t' + \eta_e}^{t' + \eta_r} d\eta' e^{-2\pi i f \eta'} W \left( t' + \eta_r - \eta' \right) = \left( \dfrac{1 - e^{-2\pi i f t}}{2\pi i f} \right) \int_{\eta_e}^{\eta_r} d\eta \ e^{-2\pi i f \eta} W\left( \eta_r - \eta \right)
\end{equation}
and
\begin{equation}
    \int_{\eta_e}^{\eta_r} d\eta e^{-2\pi i f \eta} j_l \left( 2\pi f v \left(\eta_r - \eta\right) \right) = \left( \dfrac{e^{-2\pi i f \eta_r}}{2\pi f v} \right) \int_0^{2\pi f D v} dx \ e^{i x/v} j_l\left(x\right) \,.
\end{equation}
These can be proven by a dummy variable change. The timing residual simplifies to
\begin{equation}
\begin{split}
    r\left(t,\hat{e}\right) = \ & 2 \pi \sum_A \int_{-\infty}^\infty df \int_{S^2} d\hat{k} \ \left( 1 - e^{-2\pi i f t} \right) \left( \dfrac{e^{-2\pi i f \eta_r}}{2\pi f} \right) \\
    & \ \ \ \ \times h_A\left(f, \hat{k}\right) \int_0^{2\pi f Dv} \dfrac{dx}{v} e^{ix/v} \left[ d^{ij} \varepsilon_{ij}^A \right] \sum_{lm} i^l j_l\left(x\right) Y^*_{lm}\left(\hat{k}\right) Y_{lm}\left(\hat{e}\right) \,.
\end{split}
\end{equation}
We proceed to use this to calculate the timing residual power spectrum and the overlap reduction function.

\subsection{Timing residual power spectrum and overlap reduction function}
\label{subsec:timing_and_orf}

We expand the timing residual in spherical harmonics,
\begin{equation}
    r\left(t, \hat{e}\right) = \sum_{l,m} a_{lm} Y_{lm} \left( \hat{e} \right) \,.
\end{equation}
Following the previous calculation, in the presence of a stochastic gravitational wave background comprised of a set of polarizations $A$, it can be shown that the two-point function is given by
\begin{equation}
    \langle r\left(t_a, \hat{e}_a\right) r\left(t_b, \hat{e}_b\right) \rangle = \sum_{l_1, m_1} \sum_{l_2, m_2} \langle a_{l_1 m_1} a^*_{l_2 m_2} \rangle Y_{l_1 m_1}\left( \hat{e}_a \right) Y^*_{l_2 m_2}\left( \hat{e}_b \right) \,,
\end{equation}
where
\begin{equation}
\label{eq:two_point_lm}
    \langle a_{l_1 m_1} a^*_{l_2 m_2} \rangle = \int_{-\infty}^\infty \dfrac{df}{\left(2\pi f\right)^2} \left( 1 - e^{-2\pi i f t_a} \right) \left( 1 - e^{2\pi i f t_b} \right) \sum_{A_1, A_2} \int_{S^2} d\hat{k} \ P_{A_1 A_2}\left( f, \hat{k} \right) J^{A_1}_{l_1 m_1} \left( f D_a, \hat{k} \right) J^{A_2 *}_{l_2 m_2} \left( f D_b, \hat{k} \right)
\end{equation}
and
\begin{equation}
\label{eq:Jlm_def}
    J_{lm}^A \left( fD, \hat{k} \right) = \int_0^{2\pi f D v} \dfrac{d x}{v} \ e^{i x/v} \sum_{LM} 2 \pi i^L Y^*_{LM} \left( \hat{k} \right) j_L(x) \int_{S^2} d\hat{e} \ d^{ij} \varepsilon_{ij}^A\left(\hat{k}\right) Y_{LM}\left( \hat{e} \right) Y_{lm}^*\left(\hat{e}\right) \,.
\end{equation}
In \eqref{eq:two_point_lm} and \eqref{eq:Jlm_def}, we remind that $d^{ij} = \hat{e}^i \otimes \hat{e}^j$ is the detector tensor, $\varepsilon_{ij}^A\left(\hat{k}\right)$ is the polarization basis tensor of a metric polarization $A$ propagating toward $\hat{k}$, and $P_{A_1 A_2} \left(f, \hat{k}\right)$ is the frequency space amplitude of the gravitational wave two point function, i.e.,
\begin{equation}
    \langle h_A\left(f, \hat{k}\right) h_B^* \left( f', \hat{k}'\right) \rangle = \delta \left( f - f' \right) \delta \left( \hat{k} - \hat{k}' \right) P_{AB} \left(f, \hat{k}\right) \,.
\end{equation}

Focusing on an isotropic stochastic gravitational wave background, such that $P_{AB} = \delta_{AB }P_{AA}\left(f\right)$, that is no directional dependence, the overlap reduction function measuring the angular correlation between pulsar pairs in harmonic space can be shown to be \footnote{The anisotropic case can be tackled by the replacement $Y_{00}\left(\hat{k}\right) \rightarrow Y_{lm}\left(\hat{k}\right)$ in the integral. We shall discuss this elsewhere.}
\begin{equation}
\label{eq:orf_general}
    \gamma_{ab}^A \left( \zeta, f D_i \right) = \sum_{l_1, m_1} \sum_{l_2, m_2} Y_{l_1 m_1}\left( \hat{e}_a \right) Y^*_{l_2 m_2} \left( \hat{e}_b \right) \int_{S^2} d\hat{k} \ Y_{00}\left(\hat{k}\right) J^A_{l_1 m_1} \left( f D_a, \hat{k} \right) J^{A*}_{l_2 m_2} \left( f D_b, \hat{k} \right) \,,
\end{equation}
where $\zeta$ is the angular separation between the pulsars, i.e., $\hat{e}_a \cdot \hat{e}_b = \cos \zeta$, and $Y_{00}\left(\hat{k}\right) = 1/\sqrt{4\pi}$. This is the physical observable we intend to calculate throughout this work.

\subsection{Real space formalism}
\label{subsec:realspace}

It is useful to confirm that our calculations agree with the real space formalism \cite{Chu:2021krj, NANOGrav:2021ini}. In this direction, starting with the gravitational wave \eqref{eq:gw_general} and the redshift fluctuation \eqref{eq:z_swolf}, but this time evaluating the time integrals explicitly, we find the pulsar timing residual to be
\begin{equation}
\label{eq:timing_realspace}
    r\left(t, \hat{e}\right) = \dfrac{1}{4\pi} \sum_A \int_{-\infty}^\infty \dfrac{df}{f} \int_{S^2} d\hat{k} \ \left(1 - e^{-2\pi i f t} \right) e^{-2\pi i f \eta_r} h_A\left(f, \hat{k}\right) d^{ij} \varepsilon_{ij}^A\left(\hat{k}\right) \left[ 1 - e^{2\pi i fD \left( 1 + v\hat{k}\cdot\hat{e} \right)} \right] \left( \dfrac{i}{1 + v \hat{k}\cdot\hat{e} } \right) \,.
\end{equation}
The first term in the square brackets is referred to as the `earth' term whereas the second one is the `pulsar' term, which becomes a highly oscillatory function in $\hat{k}$. Nonetheless, for the angular scales relevant in current observations, the pulsar term may be safely neglected for simplicity. Even so, we always keep the pulsar term in the formalism.

We calculate the two point function, following \cite{Chu:2021krj}, for an isotropic stochastic gravitational wave background,
\begin{equation}
\label{eq:twopoint_realspace}
\langle r\left(t_a, \hat{e}_a\right) r\left(t_b, \hat{e}_b\right) \rangle = \sum_A \int_{-\infty}^\infty df \ \left( 1 - e^{-2\pi i f t_a} \right) \left( 1 - e^{2\pi i f t_b} \right) \dfrac{P_{AA}(f)}{{2\pi^{3/2}} f^2} \times \gamma_{ab}^A\left( \zeta, f D_i \right) \,,
\end{equation}
from which we identify the overlap reduction function to be
\begin{equation}
\label{eq:orf_realspace}
    \gamma_{ab}^A\left( \zeta, f D_i \right) = \int_{S^2} \dfrac{d \hat{k}}{\sqrt{4\pi}} \ U_a \left(f D_a, \hat{k} \right) U_b^*\left(f D_b, \hat{k} \right) F_a^A \left( \hat{k} \right) F_b^{A*} \left( \hat{k} \right) \,,
\end{equation}
where
\begin{equation}
\label{eq:antenna_functions}
    F_a^A\left(\hat{k}\right) = \dfrac{d^{ij} \cdot \varepsilon_{ij}^A \left( \hat{k} \right)}{2\left( 1 + v \hat{k} \cdot \hat{e}_{{a}} \right)}
\end{equation}
and
\begin{equation}
\label{eq:Uadef}
    U_a\left(f D_a, \hat{k}\right) = 1 - e^{2 \pi i f D_a \left( 1 + v \hat{k} \cdot \hat{e}_{{a}} \right)} \,.
\end{equation}
The quantity $F_a^A\left(\hat{k}\right)$ are the so called antenna pattern functions. The quantity $H(f) \sim P_{AA}(f)$ is the one sided power spectral density of the gravitational wave background \cite{Chu:2021krj, NANOGrav:2021ini}, and is related to the fractional energy density $\Omega_{\text{GW}}(f)$ via $H(f) = \left( 3H_0^2 / (2\pi^2)\right) \times \left(\Omega_{\text{GW}}(f)/ f^3\right)$, where $H_0$ is the Hubble constant and $\Omega_\text{GW}(f) = \left( d \rho_\text{GW}/ d \ln(f) \right)/\rho_c$ for the critical energy density $\rho_c$ and gravitational wave energy density $\rho_\text{GW}$.

We utilize the real space formalism to confirm the power spectrum calculations particularly for the autocorrelation function $\gamma_{aa}^A$. This physical quantity takes in the small scale power encoded in the stochastic gravitational wave background, and so must involve at least a few thousand multipoles. By this standard, it is an incredible assessment tool for the power spectrum calculation. Our computations are presented in Table \ref{tab:GaafD100} for $fD = 100$ or $D \sim 30$ parsecs, showing the agreement between the canonical real space formalism and the power spectrum method. It is worth noting that at nonrelativistic speeds the low multipoles ($l \lesssim 100$) are enough to calculate the small scale gravitational wave power for each metric polarization.

In Appendix \ref{sec:another}, we present an alternative real space formalism that is also often considered in the literature but which starts with an explicit decomposition of the redshift fluctuation into earth and pulsar terms. It can be confirmed that this leads to the same overlap reduction function, apart from an overall factor.

\section{Tensor polarizations}
\label{sec:tensor_pols}

We derive the power spectra and overlap reduction functions for the tensor polarizations and discuss their phenomenology.

\subsection{Calculation of $J_{lm}$}
\label{subsec:Jlm_tensor}

We follow \cite{Liu:2022skj}. For convenience, we simply point the $\hat{k}$ direction to the $\hat{z}$ direction and take the magnitude to proceed.

We make use of the right and left handed complex circular polarization basis tensors:
\begin{equation}
    \varepsilon^\text{R} = \dfrac{\varepsilon^+ + i \varepsilon^\times}{\sqrt{2}} \ \ \ \ \ \text{and} \ \ \ \ \varepsilon^\text{L} = \dfrac{\varepsilon^+ - i \varepsilon^\times}{\sqrt{2}} \,.
\end{equation}
The contraction of the detector tensor with the basis tensors give
\begin{equation}
    d^{ij} \varepsilon_{ij}^\text{R, L} = \sqrt{\dfrac{16\pi}{15}} Y_{2 \pm 2} \left( \hat{e} \right) \,,
\end{equation}
where the helicity R (L) takes on $m = + 2$ ($-2$). Substituting this into \eqref{eq:Jlm_def}, and noting that
\begin{equation}
    Y_{LM}\left(\hat{k}\right) = \sqrt{\dfrac{2L+1}{4\pi}} \delta_{M0} \,
\end{equation}
since $\hat{k} = \hat{z}$, we obtain
\begin{equation}
    J_{lm}^\text{R,L} \left( fD, \hat{k} \right) = \int_0^{2\pi f D v} \dfrac{d x}{v} \ e^{i x/v} \sum_{L} 4 \pi i^L \sqrt{\dfrac{2L+1}{15}} j_L(x) \int_{S^2} d\hat{e} \ Y_{2\pm2}\left(\hat{e}\right) Y_{L0}\left( \hat{e} \right) Y_{lm}^*\left(\hat{e}\right) \,.
\end{equation}
The triple spherical harmonics integral (Appendix \ref{sec:3Y3j}) vanishes unless $m = \pm 2$ and $L = l - 2, l, l + 2$. The nonvanishing integrals are

\begin{eqnarray}
\int_{S^2} d\hat{e} \ Y_{2\pm 2}\left(\hat{e}\right) Y_{(l - 2) 0}\left(\hat{e}\right) Y^*_{l \pm 2}\left(\hat{e}\right) &=& \sqrt{\dfrac{15}{32\pi}} \left( \dfrac{(l-1)l(l+1)(l+2)}{(2l-3)(2l-1)^2(2l+1)} \right)^{1/2} \,, \\
\int_{S^2} d\hat{e} \ Y_{2\pm 2}\left(\hat{e}\right) Y_{l0}\left(\hat{e}\right) Y^*_{l \pm 2}\left(\hat{e}\right) &=& - \sqrt{\dfrac{15}{8\pi}} \left( \dfrac{(l-1)l(l+1)(l+2)}{(2l-1)^2(2l+3)^3} \right)^{1/2} \,, \\
\int_{S^2} d\hat{e} \ Y_{2\pm 2}\left(\hat{e}\right) Y_{(l + 2) 0}\left(\hat{e}\right) Y^*_{l \pm 2}\left(\hat{e}\right) &=& \sqrt{\dfrac{15}{32\pi}} \left( \dfrac{
(l-1)l(l+1)(l+2)}{(2l+1)(2l+3)^2(2l+5)} \right)^{1/2} \,.
\end{eqnarray}
Subsitutting into the last expression, we obtain
\begin{equation}
    J_{lm}^\text{R,L}\left(fD, \hat{z}\right) = -\delta_{m\pm 2} 2 \pi i^l \sqrt{ \dfrac{2l+1}{8\pi} \dfrac{(l + 2)!}{(l - 2)!} } \int_0^{2\pi fDv} \dfrac{dx}{v} \ e^{ix/v} \left( \dfrac{j_{l-2}(x)}{(2l-1)(2l+1)} + \dfrac{2j_l(x)}{(2l-1)(2l+3)} + \dfrac{j_{l+2}(x)}{(2l+1)(2l+3)} \right) \,.
\end{equation}
Then, through the recursion relation
\begin{equation}
\label{eq:bessel_id1}
    \dfrac{j_l(x)}{x} = \dfrac{j_{l-1}(x) + j_{l+1}(x)}{2l+1} \,,
\end{equation}
we are able to compactify the last expression to
\begin{equation}
    J_{lm}^\text{R,L}\left(fD, \hat{z}\right) = - \delta_{m\pm 2} \sqrt{\dfrac{2l+1}{4\pi}} \left( \sqrt{2} \pi i^l \sqrt{ \dfrac{(l + 2)!}{(l - 2)!} } \int_0^{2\pi fDv} \dfrac{dx}{v} \ e^{ix/v} \dfrac{j_l(x)}{x^2} \right) \,.
\end{equation}
We note that the factor $\sqrt{(2l + 1)/4\pi}$ corresponds to an arbitrary rotational degree of freedom. The magnitude we are interested in for an isotropic stochastic gravitational wave background is thus the one enclosed in the parenthesis in the above result. 

To generalize the result, we merely rotate the $\hat{z}$ axis into the $\hat{k} = (\theta,\phi)$ direction. This way, we obtain
\begin{equation}
    J^A_{lm}\left( fD, \hat{k} \right) = \sum_{m'} D_{m'm}^{l*} \left(-\alpha,-\theta,-\phi\right) J^A_{lm'}\left( fD, \hat{z} \right) \,,
\end{equation}
where $D_{m'm}^l(-\alpha,-\theta,-\phi)$ is the Wigner-D matrix given by
\begin{equation}
    D_{m'm}^l(-\alpha,-\theta,-\phi) = \sqrt{\dfrac{4\pi}{2l + 1}} \, _{-m'}Y_{lm}\left(\theta,\phi\right) e^{i m' \alpha} \,.
\end{equation}
Above, $\,_s Y_{lm}\left(\hat{e}\right)$ is a spin weighted spherical harmonic (Appendix \ref{sec:3Y3j}). Rotating the $\hat{z}$ to an arbitrary direction $\hat{k}$, we obtain 
\begin{equation}
\label{eq:Jlm_tensor}
    J_{lm}^\text{R,L}\left(fD, \hat{k}\right) = - _{\mp 2}Y_{lm}^* \left( \hat{k} \right) e^{\mp 2i \alpha} \left( \sqrt{2}\pi i^l \sqrt{ \dfrac{(l + 2)!}{(l - 2)!} } \int_0^{2\pi fDv} \dfrac{dx}{v} \ e^{ix/v} \dfrac{j_l(x)}{x^2} \right) \,,
\end{equation}
where the upper (lower) signs belong to R (L). We remind that the factor $e^{i m' \alpha}$ is a redundant phase owing to a remaining rotational degree of freedom about the $\hat{k}$ axis. It does not enter the observables we are interested in.

In the infinite distance limit, we may confirm that the integral admits an analytical expression:
\begin{equation}
    \int_0^{\infty} \dfrac{dx}{v} \ e^{ix/v} \dfrac{j_l(x)}{x^2} = i \sqrt{\pi } 2^{-(l+1)} (i v)^{l-2} \Gamma (l-1) \, _2\tilde{F}_1\left(\frac{l-1}{2},\frac{l}{2};l+\frac{3}{2};v^2\right) \,,
\end{equation}
where $_2\tilde{F}_1\left( a, b; c; x \right) = _2 F_1\left( a, b; c; x \right)/\Gamma(c)$ is a regularized hypergeometric function. With $v = 1$, this simplifies further to
\begin{equation}
    \int_0^\infty dx \ e^{ix} \dfrac{j_l(x)}{x^2} = 2 i^{l-1} \dfrac{(l - 2)!}{(l + 2)!} \,,
\end{equation}
which can be used to get to the Hellings-Downs correlation.

\subsection{ORF and power spectra}
\label{subsec:power_spectra_tensor}

Inserting the result into \eqref{eq:orf_general}, using the spherical harmonics addition theorem
\begin{equation}
\label{eq:addition_theorem}
    P_l \left( \hat{e}_a \cdot \hat{e}_b \right) = \dfrac{4\pi}{2l+1} \sum_{m} Y_{lm}\left(\hat{e}_a\right) Y_{lm}^*\left(\hat{e}_b\right) \,,
\end{equation}
and adding the contributions of the right and left handed helicity contributions, we obtain the overlap reduction function
\begin{equation}
    \gamma_{ab}\left( \zeta, f D_i \right) = \sum_l \dfrac{2l + 1}{4\pi} C_l P_l \left( \cos \zeta \right) \,,
\end{equation}
where $\hat{e}_a \cdot \hat{e}_b = \cos \zeta$, and the tensor power spectrum multipoles are given by
\begin{equation}
    C_l^\text{T} = \dfrac{J^\text{T}_l\left(f D_a\right) J^{\text{T}*}_l\left(f D_b\right)}{\sqrt{\pi}} \,
\end{equation}
with
\begin{equation}
    J^\text{T}_l\left(fD\right) = \sqrt{2} \pi i^l \sqrt{\dfrac{(l + 2)!}{(l - 2)!}} \int_0^{2\pi fDv} \dfrac{dx}{v} \ e^{ix/v} \dfrac{j_l(x)}{x^2} \,.
\end{equation}
For luminal tensor degrees of freedom ($v = 1$) and large pulsar distances, $fD_{a,b} \gg 1$, this reduces to the Hellings-Downs power spectrum, that is,
\begin{equation}
    C_l^\text{T} \sim \dfrac{8\pi^{3/2}}{(l + 2)(l + 1)l(l - 1)} \,.
\end{equation}

We check the power spectrum calculation compared with the real space formalism through the autocorrelation function. This requires the antenna pattern functions for the tensor $+$ and $\times$ modes which are given by
\begin{equation}
    F^+_a \left( \hat{k} = (\theta, \phi) \right) = \dfrac{\cos(2\phi) \sin^2\theta}{2 \left( 1 + v \cos \theta \right)}
\end{equation}
and
\begin{equation}
    F^\times_a \left( \hat{k} = (\theta, \phi) \right) = \dfrac{\sin(2\phi) \sin^2\theta}{2 \left( 1 + v \cos \theta \right)} \,.
\end{equation}
The tensor autocorrelation is then given by the integral
\begin{equation}
\label{eq:gammaaa_T}
    \gamma_{aa}^\text{T} = \int_0^\pi \dfrac{d\theta}{\sqrt{4\pi}} \left( \frac{2 \pi  \sin ^5 \theta  \sin ^2(\pi  fD (1 + v \cos \theta))}{(1 + v \cos \theta )^2} \right) \,. 
\end{equation}

\subsection{Phenomenology}
\label{subsec:tensor_phenomenology}

\begin{figure}[h!]
\center
\begin{adjustbox}{minipage = \linewidth, scale = 0.95}
	\subfigure[ \, tensor, $v = 0.99$, $C_2^{fD=100} = 1.82, C_2^{fD=500} = 1.82, C_2^{fD=\infty} = 1.82$ ]{
		\includegraphics[width = 0.45 \textwidth]{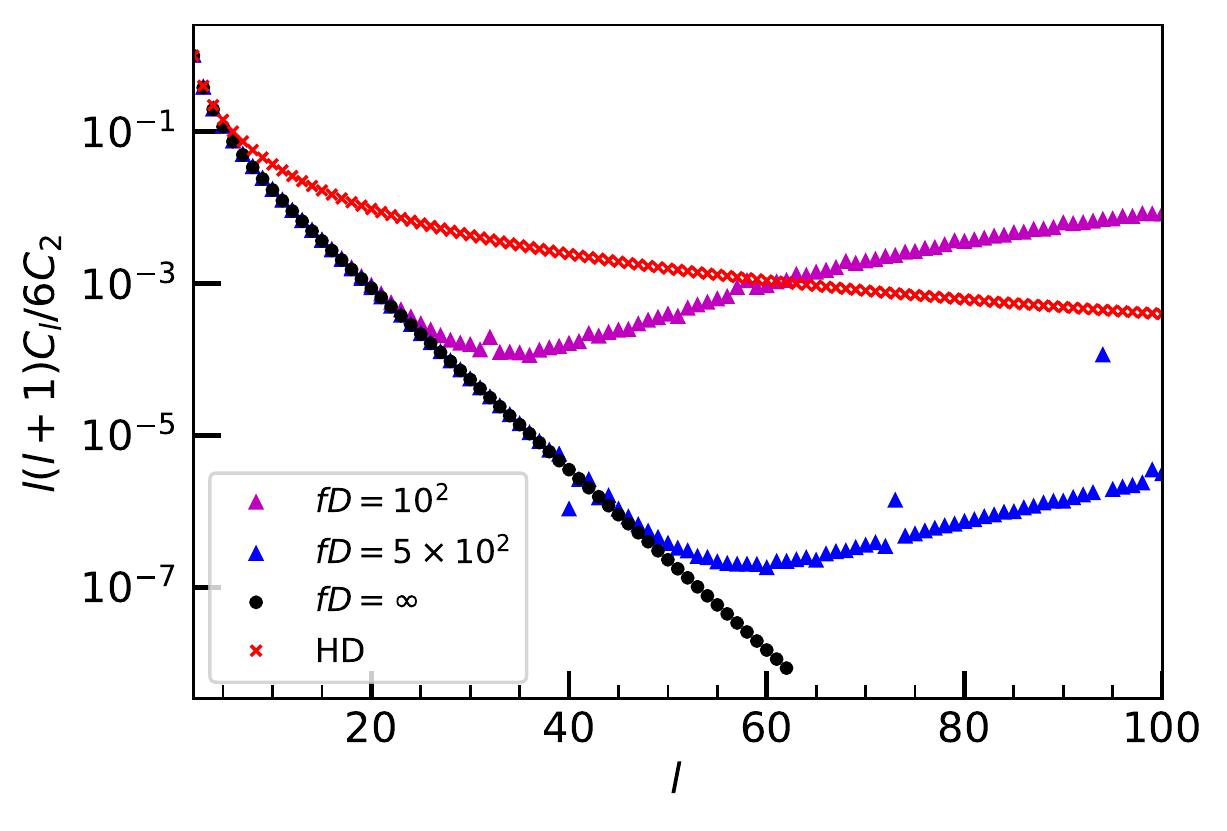}
		}
	\subfigure[ \, tensor, $v = 0.99$, $\gamma_{aa}^{fD=100} = 2.17, \gamma_{aa}^{fD=500} = 2.17, \gamma_{aa}^{fD=\infty} = 2.17$ ]{
		\includegraphics[width = 0.45 \textwidth]{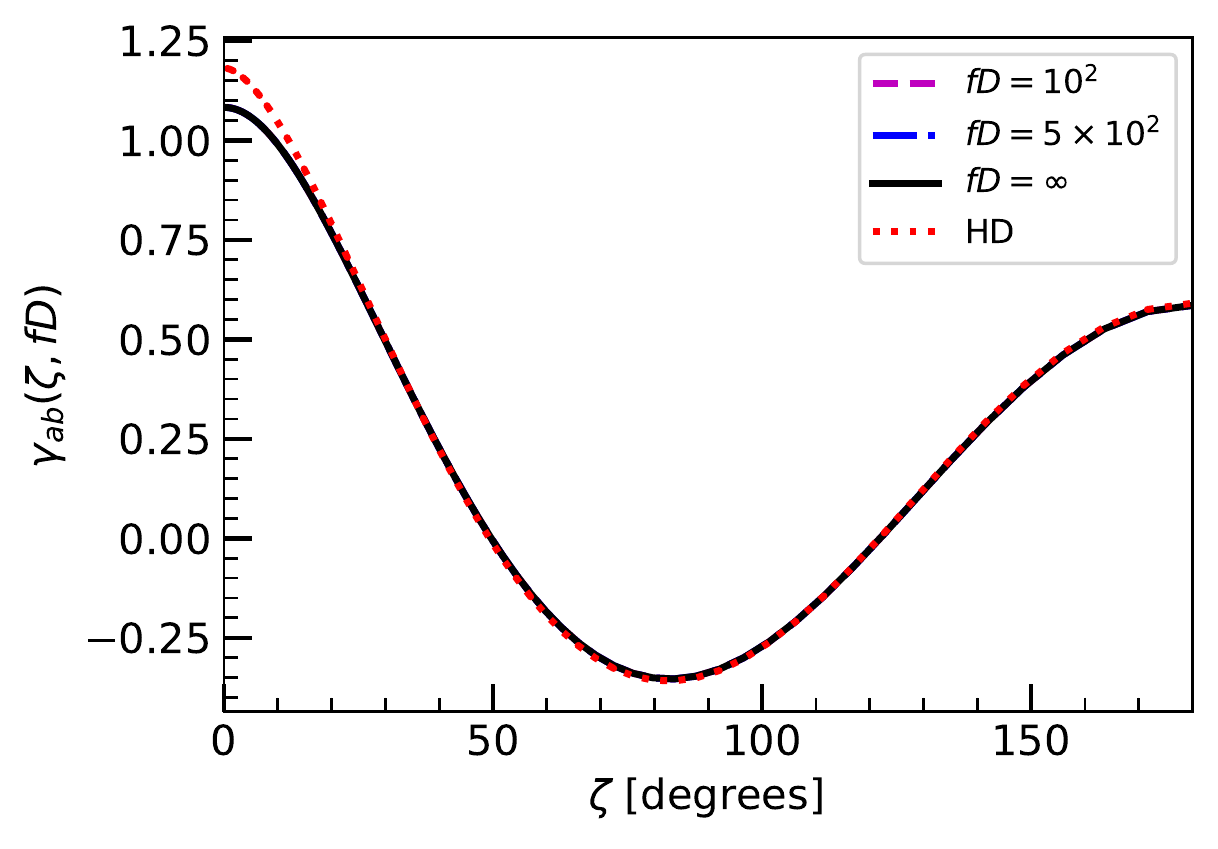}
		}
	\subfigure[ \, tensor, $v = 1/2$, $C_2^{fD=100} = 1.28, C_2^{fD=500} = 1.28, C_2^{fD=\infty} = 1.28$ ]{
		\includegraphics[width = 0.45 \textwidth]{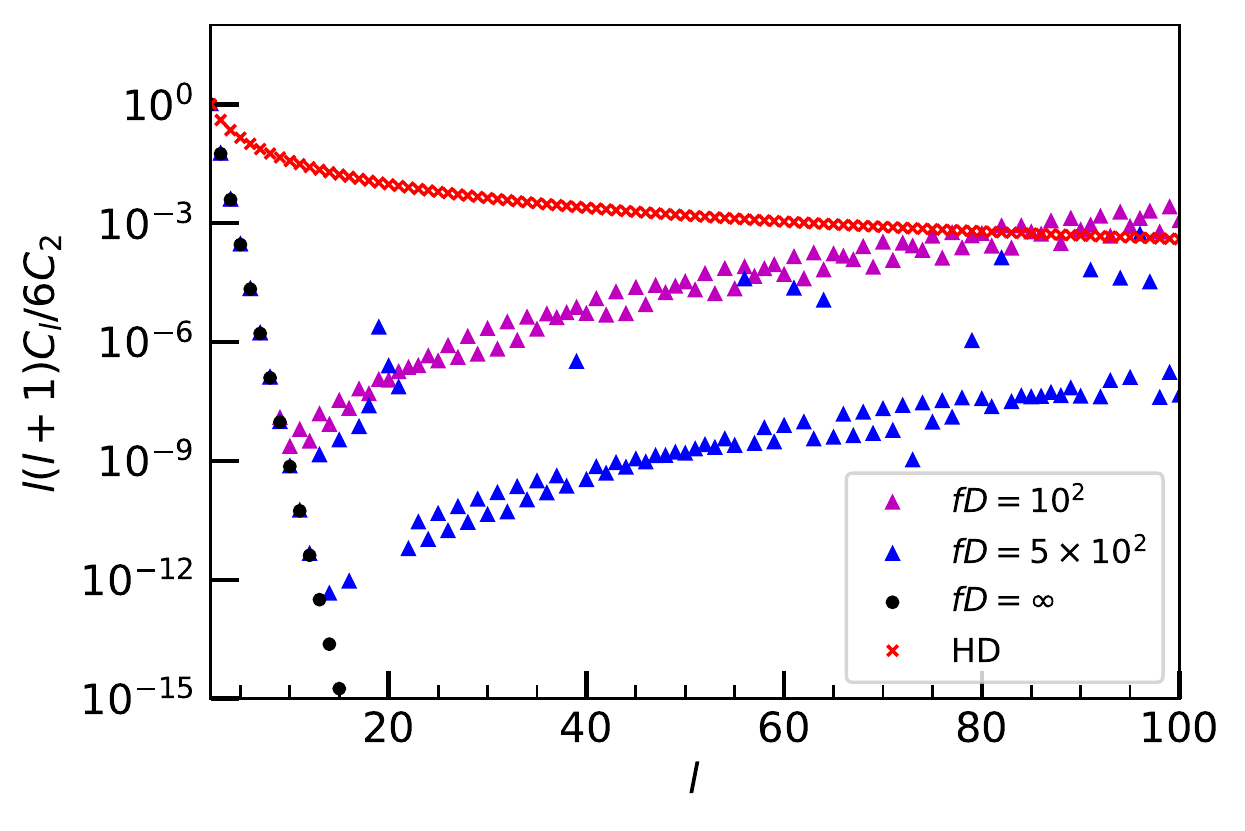}
		}
	\subfigure[ \, tensor, $v = 1/2$, $\gamma_{aa}^{fD=100} = 1.06, \gamma_{aa}^{fD=500} = 1.06, \gamma_{aa}^{fD=\infty} = 1.06$ ]{
		\includegraphics[width = 0.45 \textwidth]{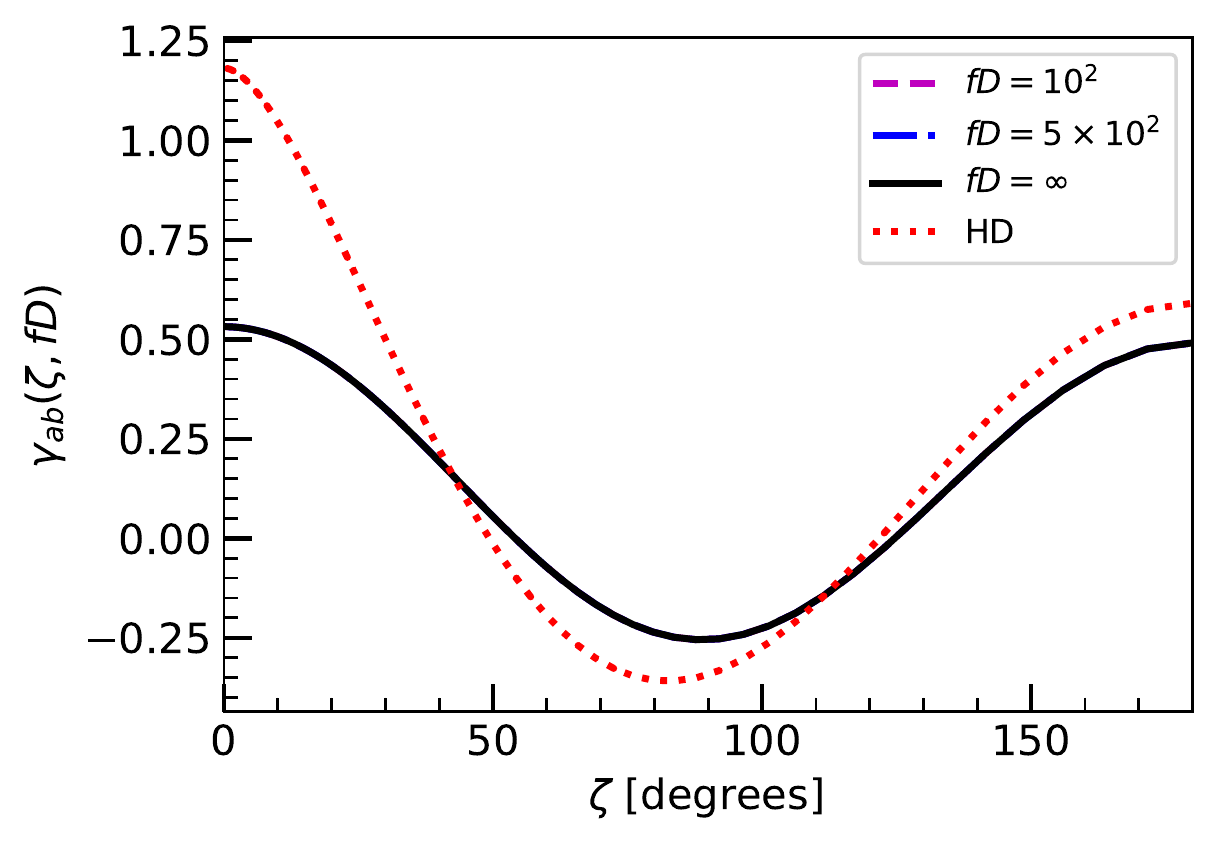}
		}
	\subfigure[ \, tensor, $v = 1/100$, $C_2^{fD=100} = 1.26, C_2^{fD=500} = 1.19, C_2^{fD=\infty} = 1.19$ ]{
		\includegraphics[width = 0.45 \textwidth]{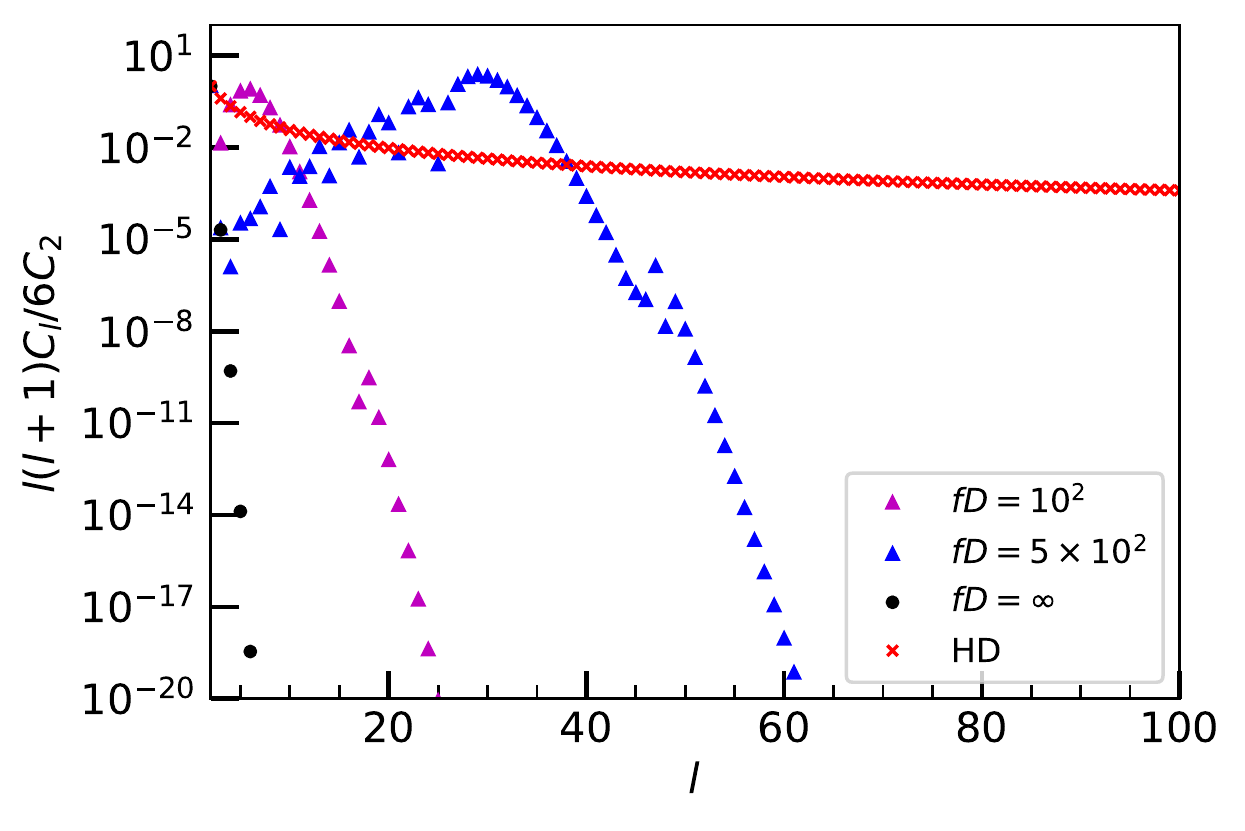}
		}
	\subfigure[ \, tensor, $v = 1/100$, $\gamma_{aa}^{fD=100} = 0.97, \gamma_{aa}^{fD=500} = 0.95, \gamma_{aa}^{fD=\infty} = 0.95$ ]{
		\includegraphics[width = 0.45 \textwidth]{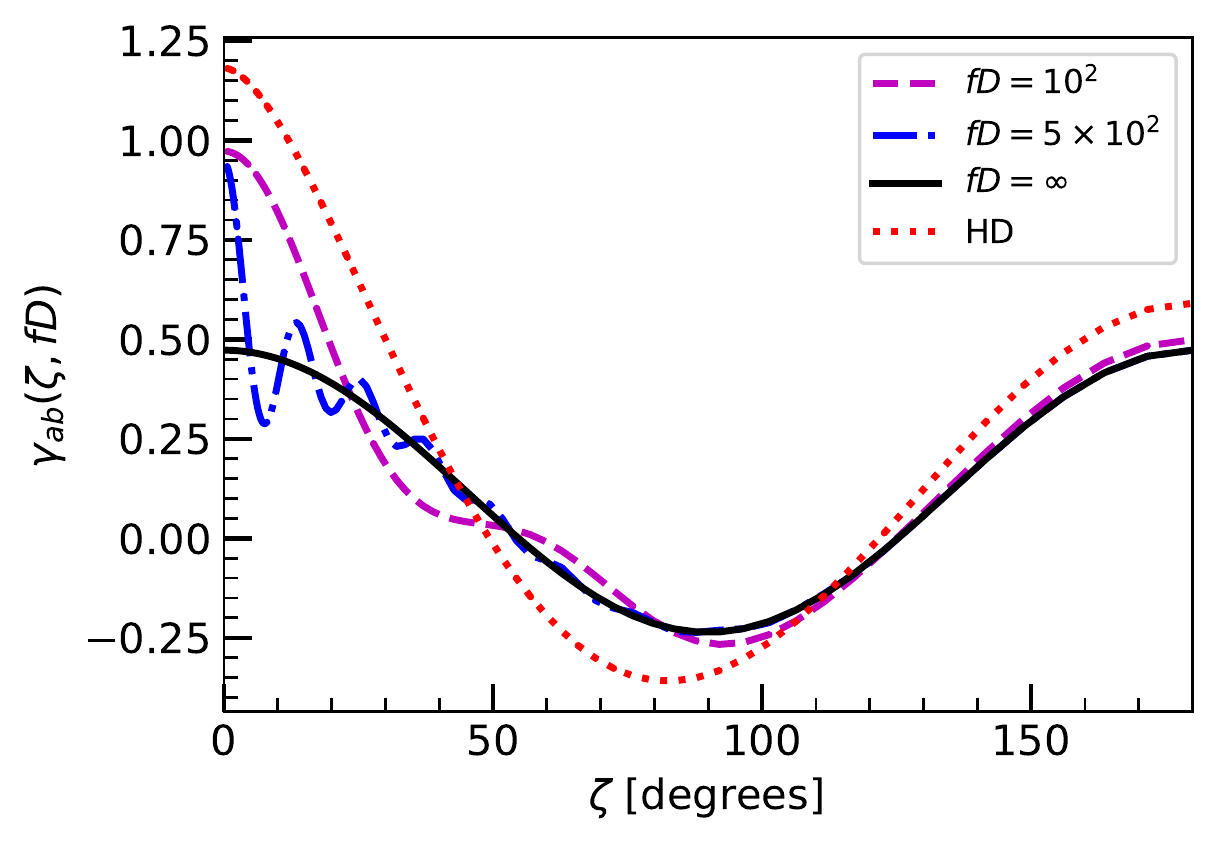}
		}
\caption{The isotropic power spectra multipoles $C_l$ and the overlap reduction functions $\gamma_{ab}\left(\zeta, fD \right)$ of the tensor polarizations with group velocities $v = 99/100$, $v = 1/2$, and $v = 1/100$. The overlap reduction functions were constructed with only the first {sixty} multipoles. The Hellings-Down correlation (HD) is shown in all plots for reference, where $C_2^{\text{HD}} = 1.86$ and the HD correlation at zero lag $\gamma_{ab}^\text{HD}(0) = 1.18$. The autocorrelation $\gamma_{aa}$'s are computed from \eqref{eq:gammaaa_T}.
}
\label{fig:ClT}
\end{adjustbox}
\end{figure}

We view the power spectrum multipoles and the ORF for various velocities and pulsar distances in Figure \ref{fig:ClT}. This is displayed together with the Hellings-Downs correlation ($v = 1$ and $fD \rightarrow \infty$) for reference.

In the near luminal ($v \sim 1$) tensor case (Figures \ref{fig:ClT}(a-b)), we reflect the small angle modifications highlighted in \cite{Ng:2021waj, Liu:2022skj} when finite pulsar distances are considered. The trend is, instead of continuously dropping as $l$ increases, the multipoles $C_l$ for finite $fD$ feature a slight growth about some $l \sim 30-50$, corresponding to an angular resolution of about $\theta \sim 3.6-6^\circ$. Also, notably, the further the pulsars are, the higher $l$ becomes to exhibit this partial sustenance. The difference is inconceivable as can be seen in the overlap reduction function where it is only the Hellings-Downs curve ($v = 1$) that is visually distinguishable from the $v = 0.99$ cases. This could be of course expected as the power spectrum in all cases are dominated by the quadrupole.

This canonical picture gradually changes as the modes go further away from the light cone. At half the speed of light (Figures \ref{fig:ClT}(c-d)), it can be seen that the finite distance modification approaches much larger scales, now with $l \sim 10-20$ being the multipole number where the $C_l$'s sustain itself. In both the finite and infinite distance case, it is most noteworthy that the spectrum becomes almost completely dominated by the quadrupole at this velocity. This manifests as an enhanced difference between the $v = 1/2$-ORF and the Hellings-Downs curve, although the finite distance cases remain indistinguishable from the infinite distance limit. This no longer holds for extreme subluminal, nonrelativistic, modes (Figure \ref{fig:ClT}(e-f)). In this limit, the infinite distance case can practically be considered a pure quadrupole. On the other hand, the finite distance cases showcase an increase at low $l$ up to some maximum power that is competitive to the quadrupole, giving a nonquadrupolar dominated power spectrum, quite like the Hellings-Downs curve but exhibiting angular oscillations beginning at the peak $l > 2$. As the peak of the power spectrum depends on the distance, the overlap reduction function also becomes distinguishable by shape depending on the distance. This clearly manifests in the overlap reduction function at this extreme subluminal velocity.

Understandably, current gravitational wave astronomy constraints in the $\sim 100$ hertz band indicate that the tensor degrees of freedom propagate on the light cone, with very little wiggle room for uncertainty. This may change in a different frequency band, as is allowed in effective field theory. However, even if it does not, we can be conservative and take the tensor modes to just be on the light cone in all frequencies, and find the analogous modifications for non tensor polarizations that could hint at modified gravity.

We interpret the enhanced small angle correlation due to the finite distance as the pulsars' perhaps interacting by some physical mechanism. Of course, such would not be case in the infinite distance limit, since the pulsars would be too far apart regardless of the size of their angular separation in the sky. This is also exhibited by the vector and scalar polarizations, as we are about to see.

\section{Vector polarizations}
\label{sec:vector_pols}

We derive the power spectra and the overlap reduction functions for the vector polarizations and discuss their phenomenology.

\subsection{Calculation of $J_{lm}$}
\label{subsec:Jlm_vector}

We simplify the calculation considerably by pointing the gravitational wave to the $\hat{z}$ direction and taking the magnitude of the result. This is sufficient for an isotropic stochastic gravitational wave background analysis.

As with the tensor polarizations, we rely on right and left handed helicity basis tensors,
\begin{equation}
    \varepsilon^\text{VR} = \dfrac{\varepsilon^x + i \varepsilon^y}{\sqrt{2}} \ \ \ \ \ \text{and} \ \ \ \ \varepsilon^\text{VL} = \dfrac{\varepsilon^x - i \varepsilon^y}{\sqrt{2}} \,,
\end{equation}
to derive the transverse vector power spectrum. The contraction of the detector tensor with the basis tensors gives
\begin{equation}
    d^{ij} \varepsilon_{ij}^\text{VR, VL} = \mp \sqrt{\dfrac{16\pi}{15}} Y_{2 \pm 1} \left( \hat{e} \right) \,,
\end{equation}
where the upper (lower) signs belong to VR (VL). The relevant spherical harmonics integrals are
\begin{equation}
    \int d\hat{e} \ Y_{21}\left(\hat{e}\right) Y_{L0}\left(\hat{e}\right) Y_{lm}\left(\hat{e}\right) = \sqrt{\dfrac{15}{2 \pi }} \dfrac{ (-l+L+1) \sqrt{l (l+1) (2 l+1) (2 L+1)} \left(-(l-1) (l+2)+L^2+L\right)}{(-l+L-2) (l+L-1) (l+L+1) (l+L+3) (l-L)! (-l+L+2)!} \,,
\end{equation}
which holds for $m = -1, l \geq 1, l - 2 \leq L \leq l + 2$ and $L + l \geq 2$ and
\begin{equation}
    \int d\hat{e} \ Y_{2-1}\left(\hat{e}\right) Y_{L0}\left(\hat{e}\right) Y_{lm}\left(\hat{e}\right) = (-1)^{L + l} \int d\hat{e} \ Y_{21}\left(\hat{e}\right) Y_{L0}\left(\hat{e}\right) Y_{lm}\left(\hat{e}\right) \,,
\end{equation}
which holds for $m = 1, l \geq 1, l - 2 \leq L \leq l + 2$ and $L + l \geq 2$. Since $L + l$ is even, the two integrals become equal except with $m = \mp 1$. We write this compactly as
\begin{equation}
\begin{split}
    \int d\hat{e} \ Y_{2\pm 1}\left(\hat{e}\right) Y_{L0}\left(\hat{e}\right) Y_{lm}\left(\hat{e}\right) = \delta_{m \mp 1}
    \bigg[ & \delta_{l1} \delta_{L1} \left( - \sqrt{\dfrac{3}{20\pi}} \right) + \delta_{l1} \delta_{L3} \sqrt{ \dfrac{9}{140\pi} } \\
    & + \Theta\left( l - 2 \right) \bigg[
    \delta_{L(l-2)} \left( - \sqrt{\dfrac{15}{2 \pi }} \dfrac{(l-1) \sqrt{l \left(4 l^3-7 l-3\right)}}{2 (2 l-3) (2 l-1) (2 l+1)} \right)
    \\
    & \phantom{ggggggggggg} + \delta_{Ll} \left( - \sqrt{\dfrac{15}{2 \pi }} \dfrac{ \sqrt{l (l+1) (2 l+1)^2}}{2 (2 l-1) (2 l+1) (2 l+3)} \right) \\
    & \phantom{ggggggggggg} + \delta_{L(l+2)} \left( \sqrt{\dfrac{15}{2 \pi }} \dfrac{ l (l+1) (l+2)}{2 (2 l+3) \sqrt{l (l+1) (2 l+1) (2 l+5)}} \right) \bigg] \bigg] \,,
\end{split}
\end{equation}
where $\Theta(x)$ is the step function. Substituting this into \eqref{eq:Jlm_def}, we get to
\begin{equation}
\begin{split}
    J_{lm}^\text{VR,VL} \left( fD, \hat{z} \right) = & \int_0^{2\pi fD v} \dfrac{dx}{v} \ e^{i x/v} \sum_{LM} 2 \pi i^L Y^*_{LM}\left( \hat{k} \right) j_L(x) \int_{S^2} d\hat{e} \ \left( d^{ij} \varepsilon_{ij}^\text{VR, VL} \left( \hat{k} \right) \right) Y_{LM}\left(\hat{e}\right) Y_{lm}^*\left(\hat{e}\right) \\
    = & \mp \sqrt{ \dfrac{16\pi}{15} } \int_0^{2\pi f D v} \dfrac{dx}{v} \ e^{ix/v} \sum_L 2 \pi i^L \sqrt{\dfrac{2L+1}{4\pi}} j_L(x) \int_{S^2} d\hat{e} \ Y_{2\pm 1} \left(\hat{e}\right) Y_{L0}\left(\hat{e}\right) Y_{lm}^* \left(\hat{e}\right) \\
    = & \mp \sqrt{ \dfrac{16\pi}{15} } \int_0^{2\pi f D v} \dfrac{dx}{v} \ e^{ix/v} \delta_{m \pm 1} \bigg[ - \delta_{l1} \dfrac{3i}{2 \sqrt{5}} \left(j_1(x) + j_3(x)\right) \\
    & + \Theta(l - 2) \sqrt{\dfrac{15}{2}} \dfrac{i^l}{2} \bigg[
    \dfrac{(l-1) \sqrt{l (l+1) (2 l+1)} j_{l-2}(x)}{4 l^2-1}
    \\
    & \phantom {ggggggggggggggggg} -\frac{2 \sqrt{l (l+1) (2 l+1)} j_l(x)}{8 l (l+1)-6} -\dfrac{2 (l+2) \sqrt{l (l+1) (2 l+1)} j_{l+2}(x)}{8 l (l+2)+6} \bigg] \bigg] \,.
\end{split}
\end{equation}
The last expression simplifies to
\begin{equation}
\begin{split}
    J_{lm}^\text{VR,VL} \left( fD, \hat{z} \right) = \mp \sqrt{ \dfrac{16\pi}{15} } \delta_{m \pm 1} \bigg[ & - \delta_{l1} \dfrac{3i}{2 \sqrt{5}} \int_0^{2\pi f D v} \dfrac{dx}{v} \ e^{ix/v} \left(j_1(x) + j_3(x)\right) \\
    & + \Theta(l - 2) \sqrt{\dfrac{15}{2}} \dfrac{i^l}{2} \sqrt{l (l + 1)(2l + 1)} \int_0^{2\pi f D v} \dfrac{dx}{v} \ e^{ix/v} \dfrac{d}{dx} \left( \dfrac{j_l(x)}{x} \right) \bigg] \,.
\end{split}
\end{equation}
Now, the $l = 1$ piece above can be continued to give the same expression as the $l \geq 2$ pieces. We therefore have
\begin{equation}
\begin{split}
    J_{(l \geq 1)m}^\text{VR,VL} \left( fD, \hat{z} \right) = \mp \delta_{m \pm 1} \sqrt{\dfrac{2l + 1}{4\pi}} \left( 2 \sqrt{2} \pi i^l \sqrt{l (l + 1)} \int_0^{2\pi f D v} \dfrac{dx}{v} \ e^{ix/v} \dfrac{d}{dx} \left( \dfrac{j_l(x)}{x} \right) \right) \,.
\end{split}
\end{equation}
We take the magnitude above aside from the rotation factor $\sqrt{(2l + 1)/4\pi}$ to compute the overlap reduction function of an isotropic gravitational wave background.

We rotate the $\hat{z}$ direction to an arbitrary $\hat{k}$. This leads to
\begin{equation}
\begin{split}
    J_{(l \geq 1)m}^\text{VR,VL} \left( fD, \hat{k} \right) = \mp \, _{\mp 1} Y_{lm}^*\left(\hat{k}\right) e^{\mp i \alpha} \left( 2\sqrt{2} \pi i^l \sqrt{l (l + 1)} \int_0^{2\pi f D v} \dfrac{dx}{v} \ e^{ix/v} \dfrac{d}{dx} \left( \dfrac{j_l(x)}{x} \right) \right) \,.
\end{split}
\end{equation}
By integration by parts, it is useful to note that the integral can be written as
\begin{equation}
\label{eq:Jlm_vec}
    \int_0^{2\pi f D v} \dfrac{dx}{v} \ e^{ix/v} \dfrac{d}{dx} \left( \dfrac{j_l(x)}{x} \right) = - \dfrac{i}{v} \int_0^{2\pi f D v} \dfrac{dx}{v} \ e^{ix/v} \dfrac{j_l(x)}{x} + \dfrac{e^{2\pi i fD}}{v} \dfrac{j_l\left(2\pi f D v\right)}{2\pi f D v} - \dfrac{\sqrt{\pi } 2^{-(l + 1)}}{v \Gamma \left(l+(3/2)\right)} \epsilon^{l - 1}|_{\epsilon \rightarrow 0^+} \,.
\end{equation}
The phase factor $e^{\pm i \alpha}$ corresponds to an arbitrary rotational degree of freedom along the $\hat{k}$ direction. This drops out in the physical observables of interest in this work. We also put attention to the boundary terms (second and third terms in the right hand side) in \eqref{eq:Jlm_vec}. The first one comes from the finite distance modification, as is clear this vanishes when $fD \rightarrow \infty$. The second one vanishes for $l > 1$ but reduces to a constant ($\sim 1/v$) for the dipole $l = 1$.

We note that the integral admits an analytical expression for the infinite distance case:
\begin{equation}
    \int_0^{\infty} \dfrac{dx}{v} \ e^{ix/v} \dfrac{j_l(x)}{x} = \sqrt{\pi } 2^{-(l+1)} i^l v^{l - 1} \Gamma (l) \, _2\tilde{F}_1\left(\frac{l}{2},\frac{l+1}{2};l+\frac{3}{2};v^2\right) \,.
\end{equation}
In the luminal limit, this further simplifies to
\begin{equation}
    \int_0^{\infty} dx \ e^{ix} \dfrac{j_l(x)}{x} = i^l \dfrac{(l - 1)!}{(l + 1)!} \,,
\end{equation}
which can be used to derive the analogous Hellings-Downs correlation ($v = 1$ and $fD \rightarrow \infty$) for the vector modes.

\subsection{ORF and power spectra}
\label{subsec:power_spectra_vector}

Inserting the result to \eqref{eq:orf_general}, using the addition theorem, and taking in the contributions from the left and right handed helicity vector polarizations, we obtain the overlap reduction function for a vector sourced isotropic stochastic gravitational wave background:
\begin{equation}
    \gamma_{ab} \left( \zeta, f D_i \right) = \sum_l \dfrac{2l + 1}{4\pi} C_l P_l \left( \cos \zeta \right) \,,
\end{equation}
where the vector power spectrum is
\begin{equation}
    C_l^\text{V} = \dfrac{J^\text{V}_l\left(f D_a\right) J^{\text{V}*}_l\left(f D_b\right)}{\sqrt{\pi}} \,
\end{equation}
with the function $J^\text{V}_l(fD)$ given by
\begin{equation}
    J^\text{V}_l\left(fD\right) = 2 \sqrt{2} \pi i^l \sqrt{l(l+1)} \int_0^{2\pi fDv} \dfrac{dx}{v} \ e^{ix/v} \dfrac{d}{dx} \left( \dfrac{j_l(x)}{x} \right) \,.
\end{equation}

As with the tensor, we validate the power spectrum calculation by comparing it with the real space formalism through the calculation of the autocorrelation function. The antenna pattern functions for the vector $x$ and $y$ modes are
\begin{equation}
    F^x_a \left( \hat{k} = (\theta, \phi) \right) = \dfrac{\sin(2\theta) \cos \phi}{2 \left( 1 + v \cos \theta \right)}
\end{equation}
and
\begin{equation}
    F^y_a \left( \hat{k} = (\theta, \phi) \right) = \dfrac{\sin(2\theta) \sin \phi}{2 \left( 1 + v \cos \theta \right)} \,.
\end{equation}
The vector autocorrelation reduces to the integral
\begin{equation}
\label{eq:gammaaa_V}
    \gamma_{aa}^\text{V} = \int_0^\pi \dfrac{d\theta}{\sqrt{4\pi}} \left( \frac{8 \pi  \sin ^3 \theta \cos ^2 \theta \sin ^2(\pi fD (1 + v \cos \theta ))}{(1 + v \cos \theta )^2} \right) \,. 
\end{equation}

\subsection{Phenomenology}
\label{subsec:vector_phenomenology}

\begin{figure}[h!]
\center
\begin{adjustbox}{minipage = \linewidth, scale = 0.95}
	\subfigure[ \, vector, $v = 1$, $C_2^{fD=100} = 6.24, C_2^{fD=500} = 6.23, C_2^{fD=\infty} = 6.23$ ]{
		\includegraphics[width = 0.45 \textwidth]{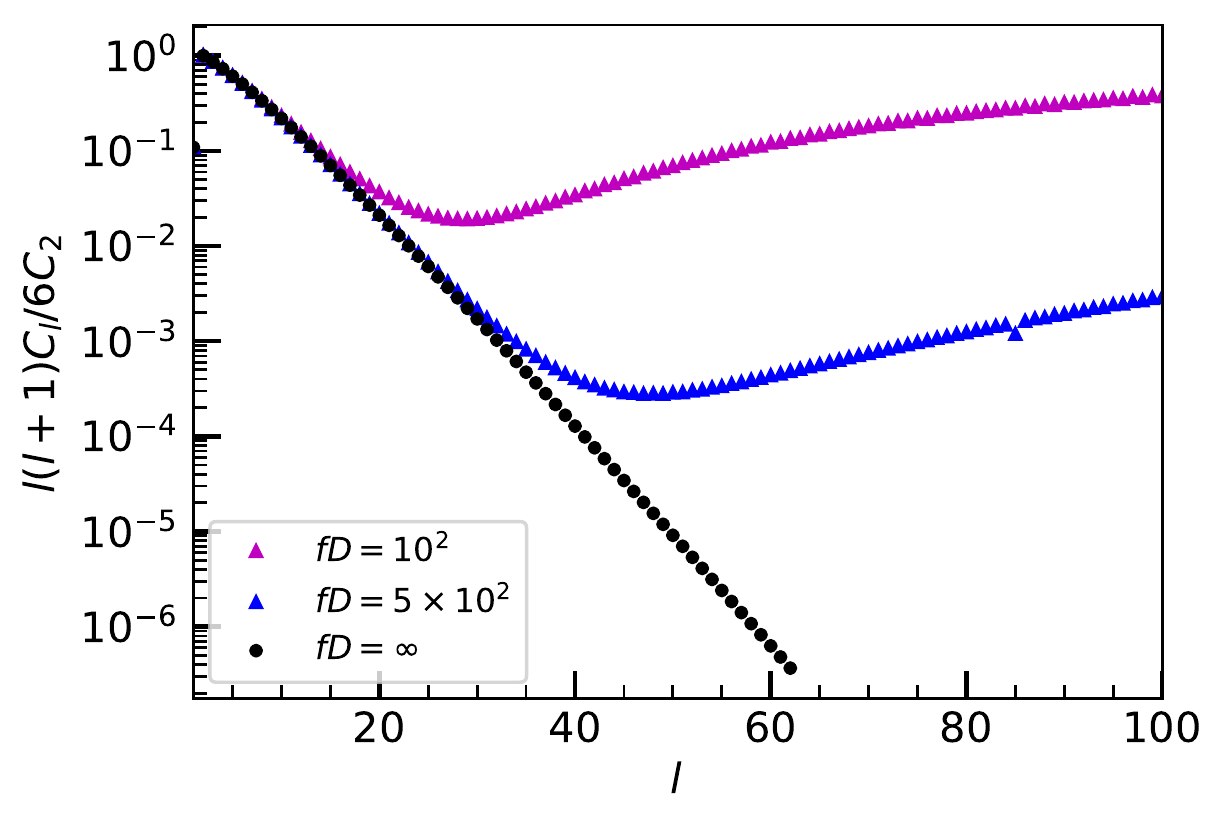}
		}
	\subfigure[ \, vector, $v = 1$, $\gamma_{aa}^{fD=100} = 15.7, \gamma_{aa}^{fD=500} = 15.6, \gamma_{aa}^{fD=\infty} = 15.5$ ]{
		\includegraphics[width = 0.45 \textwidth]{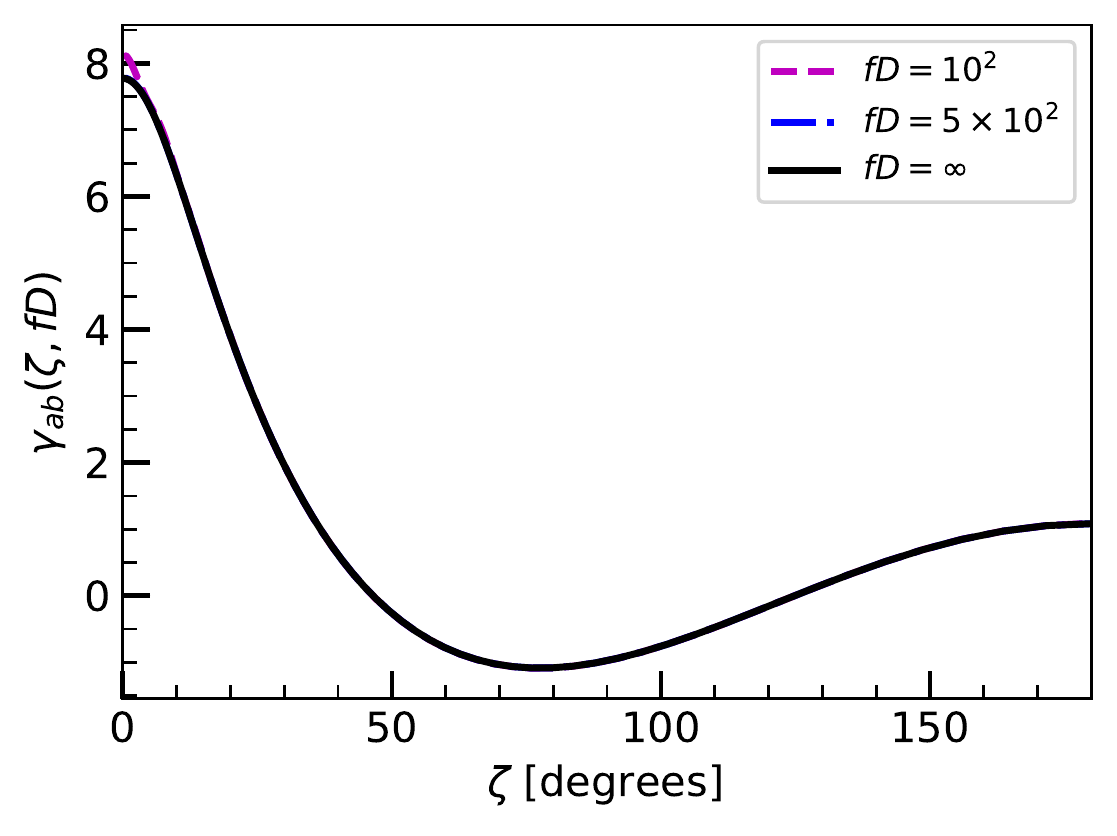}
		}
	\subfigure[ \, vector, $v = 1/2$, $C_2^{fD=100} = 1.50, C_2^{fD=500} = 1.50, C_2^{fD=\infty} = 1.50$ ]{
		\includegraphics[width = 0.45 \textwidth]{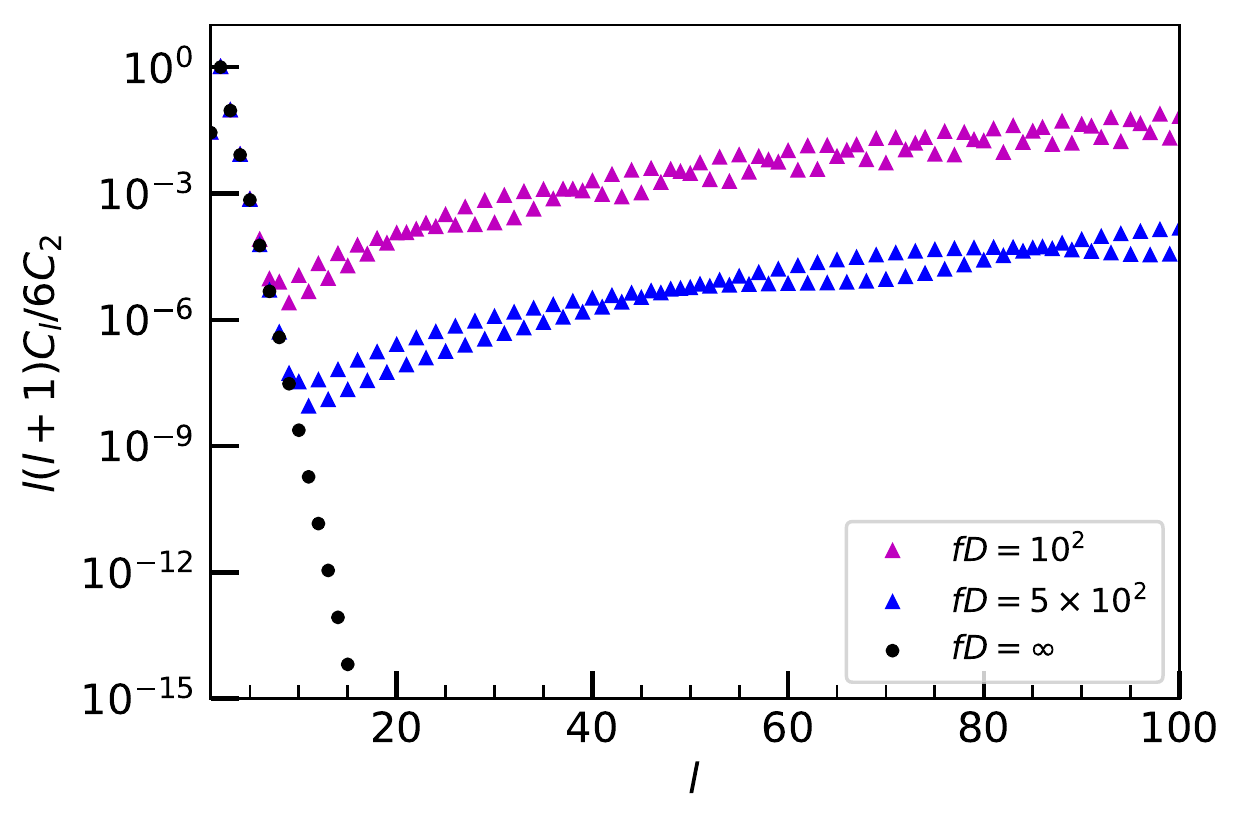}
		}
	\subfigure[ \, vector, $v = 1/2$, $\gamma_{aa}^{fD=100} = 1.34, \gamma_{aa}^{fD=500} = 1.34, \gamma_{aa}^{fD=\infty} = 1.34$ ]{
		\includegraphics[width = 0.45 \textwidth]{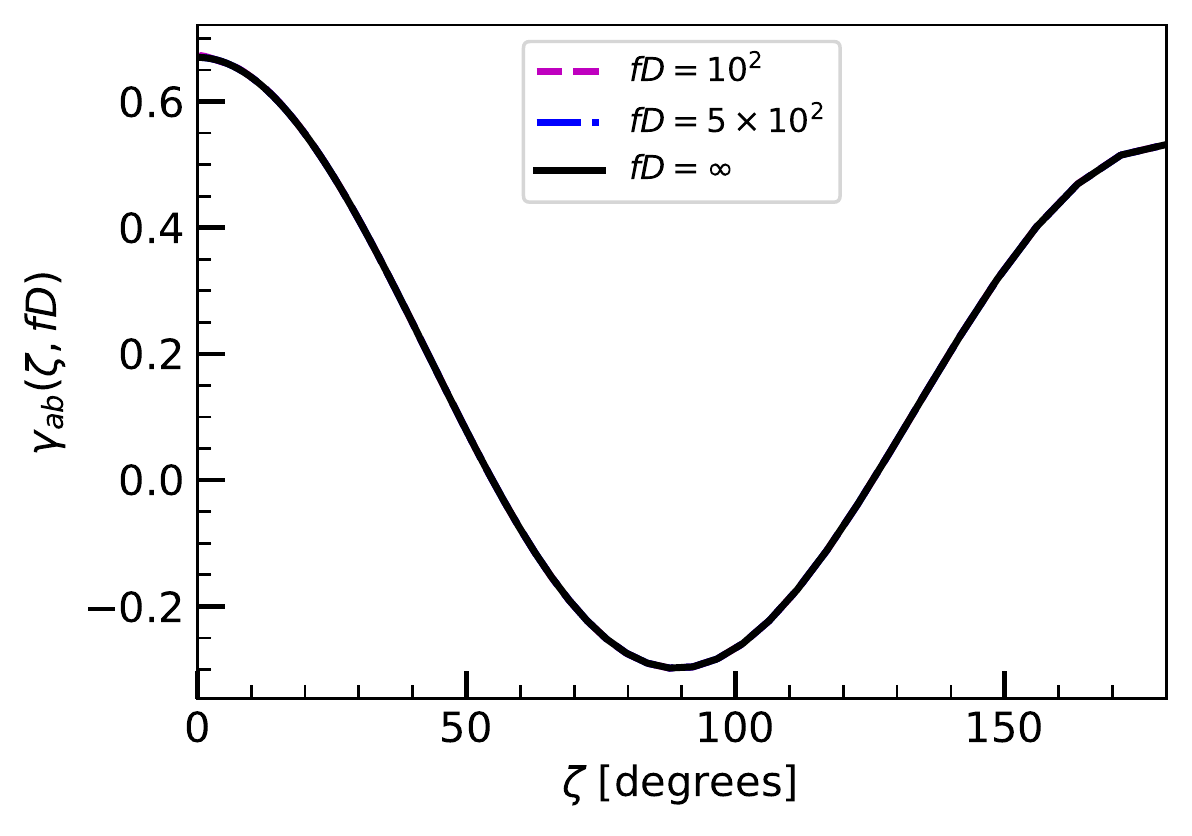}
		}
	\subfigure[ \, vector, $v = 1/100$, $C_2^{fD=100} = 1.90, C_2^{fD=500} = 1.22, C_2^{fD=\infty} = 1.19$ ]{
		\includegraphics[width = 0.45 \textwidth]{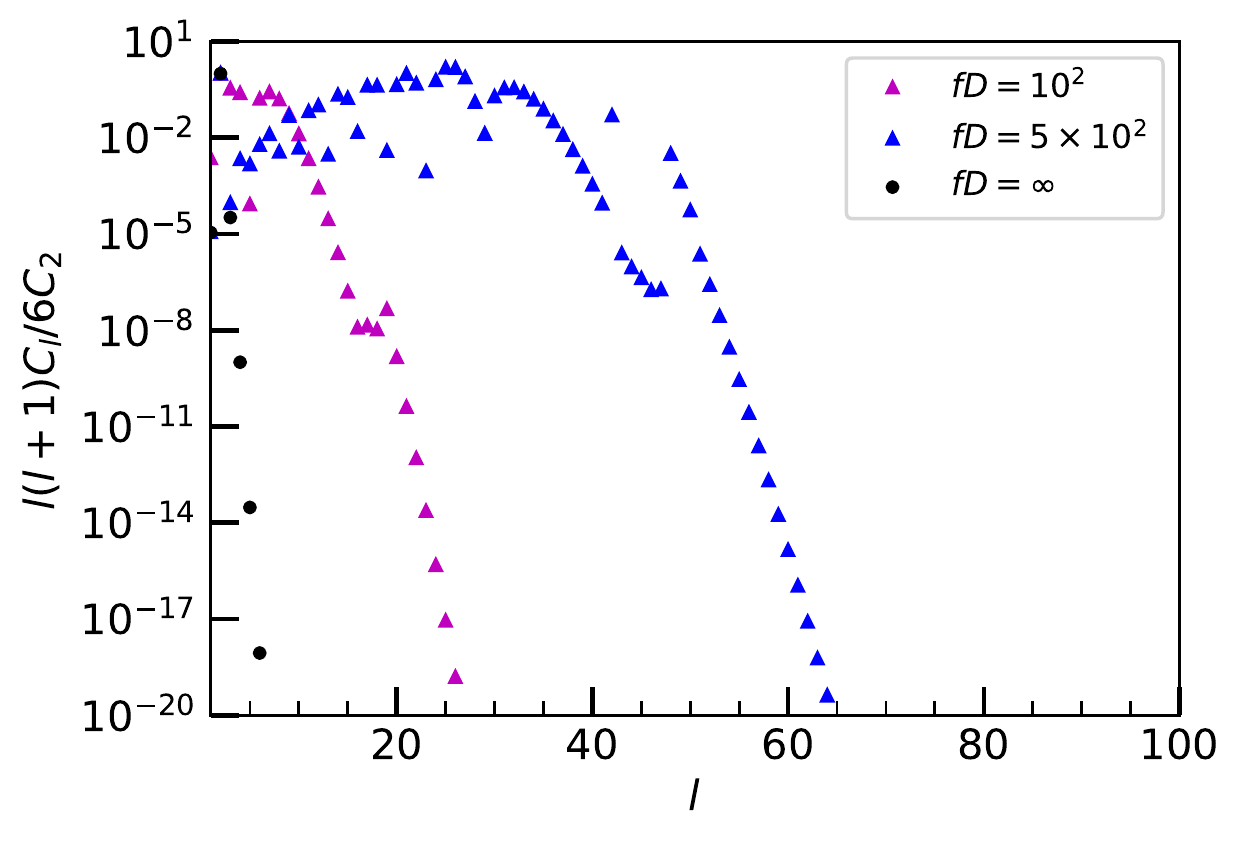}
		}
	\subfigure[ \, vector, $v = 1/100$, $\gamma_{aa}^{fD=100} = 1.20, \gamma_{aa}^{fD=500} = 0.96, \gamma_{aa}^{fD=\infty} = 0.95$ ]{
		\includegraphics[width = 0.45 \textwidth]{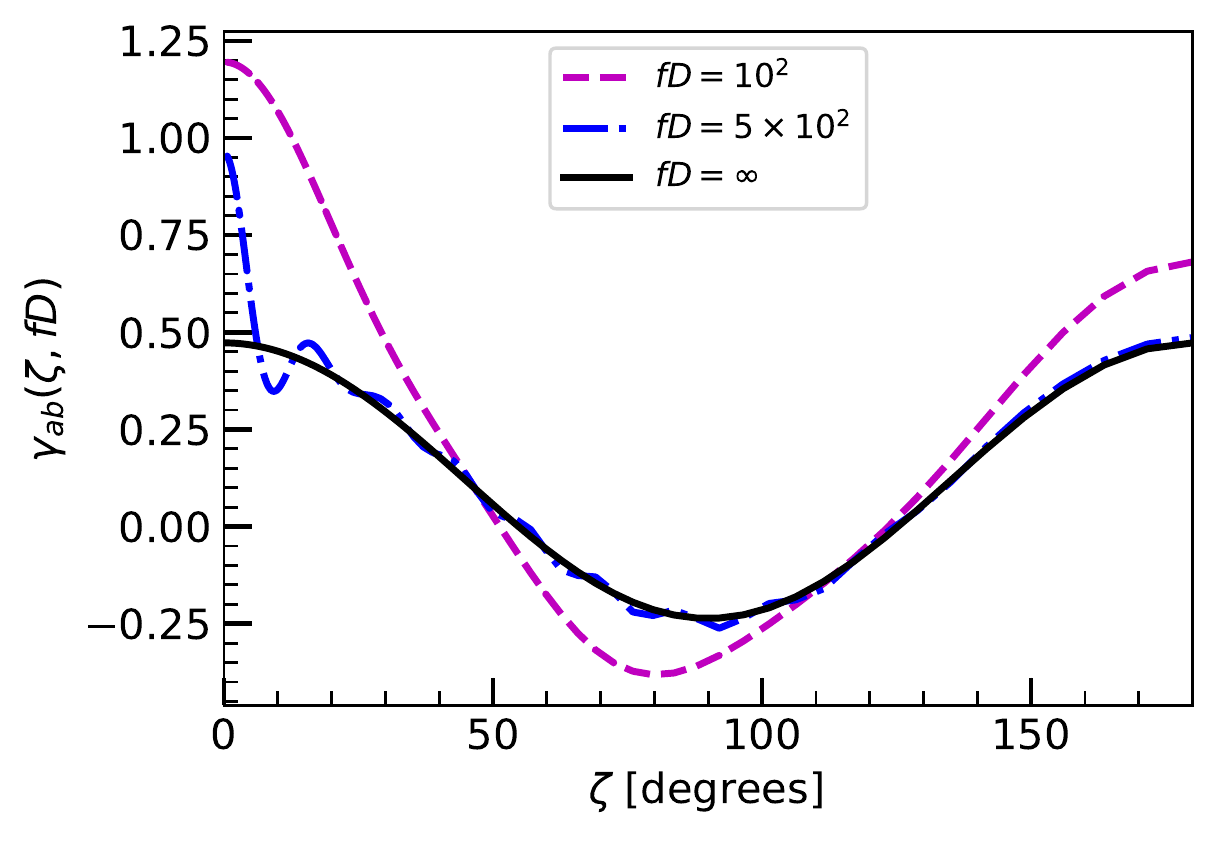}
		}
\caption{The isotropic power spectra multipoles $C_l$ and the overlap reduction functions $\gamma_{ab}\left(\zeta, fD \right)$ of the vector polarizations with group velocities $v = 99/100$, $v = 1/2$, and $v = 1/100$. The overlap reduction functions were constructed with only the first {sixty} multipoles. The autocorrelation $\gamma_{aa}$'s are computed from \eqref{eq:gammaaa_V}.
}
\label{fig:ClV}
\end{adjustbox}
\end{figure}

Figure \ref{fig:ClV} presents the power spectra multipoles and the corresponding ORFs for vector polarizations of various velocities and distances.

For the near luminal vector (Figures \ref{fig:ClV}(a-b)), it can be seen that throughout the vector power spectrum drops differently compared with the Hellings-Downs correlation. It is noteworthy that the dipolar power is also suppressed compared to the quadrupole and the succeeding multipoles such as the octupole and so on, regardless of the pulsars' distances. This shows up as a difference in the {ORF} at large angles for the vector induced correlation compared with the Hellings-Downs curve. As with the tensor modes, the power spectrum starts to sustain itself at some $l \sim 30 - 50$, corresponding to an angular resolution $3.6-6^\circ$. However, this difference is realizable only at small angles, irrelevant for the current pulsar timing array data, which reflects as the indistinguishability of the {ORFs} for various pulsar distances and the infinite distance limit. We realize an angular dependence due to the pulsar distance to be more pronounced for subluminal velocities.

At half the speed of light (Figures \ref{fig:ClV}(c-d)), we find that the dipole becomes more relevant, but still supressed compared to the quadrupole and the octupole. This is reflected in the power spectrum and the {ORF} which appears to be shaped more like the Hellings-Downs curve compared with the luminal vector case. The difference can be attributed due to the quadrupole, being too dominant in the vector power spectrum. This time, the multipole number at which higher modes start to sustain themselves becomes lower, $l \sim 5-10$, making the small angle departure more realizable provided sufficient sensitivity. Yet, as the power drop per multipole becomes steeper for subluminal velocity, it remains to distinguish between the finite and infinite pulsar distance cases. This picture drastically changes for the nonrelativistic vector (Figures \ref{fig:ClV}(e-f)), where low multipoles other than the dipole and the quadrupole contribute significantly to the power spectrum. In particular, in Figure \ref{fig:ClV}(f), the {ORF} for the vector with $fD = 100$ becomes strikingly similar to the Hellings-Downs curve, while for the $fD = 500$ and infinite distance cases, it is not. This can also be realized in the vector power spectrum multipoles for $fD = 100$ where it can be seen that the dipole is suppressed at this extreme subluminal velocity while the low multipoles beginning with the quadrupole and the octupole follow the trend of the Hellings-Downs correlation. The picture changes further for more distant pulsars, at $fD = 500$, in which case the higher multipoles $l \sim 20-30$ can be seen to even be as significant as the quadrupole. This changes the shape of the vector {ORF} at all angles while still being dominated by the quadrupole, as it manifests visually.

The above results tease a degeneracy in the tensor and vector degrees of freedom, particularly with the luminal tensor and nonrelativistic vector, in the {ORF} and the present data set. Nonetheless, this can be settled by resolving small angular separations, which may be realizable in upcoming pulsar timing array missions.

\section{Scalar polarizations}
\label{sec:scalar_pols_finite_dist}

We derive the power spectra and the overlap reduction functions for the scalar polarizations and study their phenomenology.

\subsection{Calculation of $J_{lm}$}
\label{subsec:calcu_Jlm}

As we did with the tensor and vector cases, to calculate the overlap reduction function for an isotropic gravitational wave background, we simply choose $\hat{k} = \hat{z}$ direction and pick up the magnitude to take in \eqref{eq:orf_general}.

The contraction $d^{ij} \varepsilon_{ij}$ of the detector tensor and the polarization basis for the scalar transverse and scalar longitudinal modes becomes
\begin{equation}
\label{eq:deps_ST}
    d^{ij} \varepsilon_{ij}^{\text{ST}} = \sin^2 \theta
\end{equation}
and
\begin{equation}
\label{eq:deps_SL}
    d^{ij} \varepsilon_{ij}^{\text{SL}} = \sqrt{2} \cos^2 \theta \,.
\end{equation}
Since these appear in \eqref{eq:Jlm_def} together with two more spherical harmonics $Y_{lm}\left(\hat{e}\right)$ in an integral, it is useful to express the above contractions as a spherical harmonic series:
\begin{equation}
    d^{ij} \varepsilon_{ij}^{\text{ST}} = \dfrac{4\sqrt{\pi}}{3} Y_{00}\left(\hat{e}\right) - \dfrac{4}{3} \sqrt{\dfrac{\pi}{5}} Y_{20}\left(\hat{e}\right)
\end{equation}
and
\begin{equation}
    d^{ij} \varepsilon_{ij}^{\text{SL}} = \sqrt{2} \left( \dfrac{2\sqrt{\pi}}{3} Y_{00}\left(\hat{e}\right) + \dfrac{4}{3} \sqrt{\dfrac{\pi}{5}} Y_{20}\left(\hat{e}\right) \right) \,.
\end{equation}
We perform the summation and integration in \eqref{eq:Jlm_def} for each of the scalar polarizations. After simplification, a rotation is then acted on the result to generalize it to a gravitational wave propagating in an arbitrary direction $\hat{k}$.

We start with the scalar transverse polarization. The only term which survives the sum over ${L,M}$ in \eqref{eq:Jlm_def} is $M = 0$. Consequently, the three spherical harmonics integrals we need are
\begin{equation}
\label{eq:IntY3_Scalar1}
    \int d \hat{e} \ Y_{00}\left(\hat{e}\right) Y_{L0}\left(\hat{e}\right) Y_{l m}\left(\hat{e}\right) = \dfrac{\delta_{m0}\delta_{lL}}{\sqrt{4\pi}} 
\end{equation}
and
\begin{equation}
\label{eq:IntY3_Scalar2}
    \int d \hat{e} \ Y_{20}\left(\hat{e}\right) Y_{L0}\left(\hat{e}\right) Y_{l m}\left(\hat{e}\right) = \dfrac{ \delta_{m0} }{2} \sqrt{\dfrac{5}{\pi}} \dfrac{(L-l +1)^2 (L+l ) (L+l +2) \sqrt{(2 L+1) (2 l +1)} \Gamma (-L+l +3)}{(-L+l +2)^2 (L+l -1) (L+l +1) (L+l +3) \Gamma\left(l -L+1\right)^2 \Gamma (L-l +3)} \,,
\end{equation}
where \eqref{eq:IntY3_Scalar2} holds provided $l - 2 \leq L \leq l + 2$ and $L + l \geq 2$; otherwise, it is zero. Now, in performing the sum over $L$, we note that $l_1 + l_2 + l_3$ in the Wigner-$3j$ symbol must be an even integer for our purposes since $m_1 = m_2 = 0$ (thus consequently setting up $m_3 = 0$). This leaves three terms corresponding to $L = l - 2, l, l + 2$. Also, from the $L = l + 2$ contribution, we may pull out $l = 0, 1$ terms. Likewise, from the $L = l$ contribution, we may pull out $l = 1$. In this way, we can add the terms for $l \geq 2$ coming from all $L = l - 2, l, l + 2$ terms. This way, we are able to write down the last integral as
\begin{equation}
\label{eq:IntY3_Scalar2b}
    \int d \hat{e} \ Y_{20}\left(\hat{e}\right) Y_{L0}\left(\hat{e}\right) Y_{l m}\left(\hat{e}\right) =
    \begin{cases}
    \dfrac{3\delta_{m0}}{4} \sqrt{\dfrac{5}{\pi }} \dfrac{ (l-1) l}{ \sqrt{2 l-3} (2 l-1) \sqrt{2 l+1}} &, \ \ \ \ L = l-2 , l \geq 2 \phantom{\dfrac{\dfrac{1}{2}}{\dfrac{1}{2}}} \\
    \delta_{m0} \sqrt{\dfrac{5}{\pi }} \dfrac{ l (l+1)}{2(2l - 1)(2l+3)} &, \ \ \ \ L = l , l \geq 1 \phantom{\dfrac{\dfrac{1}{2}}{\dfrac{1}{2}}} \\
    \dfrac{3\delta_{m0}}{4}\sqrt{\dfrac{5}{\pi }} \dfrac{ (l+1) (l+2)}{(2 l+3) \sqrt{(2 l+1) (2 l+5)}} &, \ \ \ \ L = l+2 , l \geq 0 \phantom{\dfrac{\dfrac{1}{2}}{\dfrac{1}{2}}} \,.
    \end{cases} 
\end{equation}

Using the above identities, and carefully performing the sum over $L$, we get to
\begin{equation}
\label{eq:J00_STc}
\begin{split}
    J_{lm}^\text{ST} \left( fD, \hat{z} \right)
    = \ & \delta_{m0} \delta_{l0} \dfrac{2\sqrt{ \pi}}{3} \int_0^{2\pi fDv} \dfrac{d x}{v} \ e^{i x/v} \left( j_0(x) + j_2(x) \right) + \delta_{m0}\delta_{l1} \dfrac{2 \sqrt{3\pi} i}{5}\int_0^{2\pi fDv} \dfrac{d x}{v} \ e^{i x/v} \left( j_1(x) + j_3(x) \right) \\
    & \ \ - \delta_{m0} \Theta\left(l - 2\right) 4\pi i^l \sqrt{\dfrac{2l+1}{4\pi}} \int_0^{2\pi fDv} \dfrac{d x}{v} \ e^{i x/v} \left[ \dfrac{d}{dx} \left( \dfrac{j_l(x)}{x} \right) -\dfrac{(l-1)(l+2)}{2} \dfrac{j_l(x)}{x^2} \right] \,.
\end{split}
\end{equation}
By using the spherical Bessel function differential equation,
\begin{equation}
x^2 j_l''(x) + 2xj_l'(x) + \left( x^2 - l(l+1) \right) j_l(x) = 0 \,,    
\end{equation}
and \eqref{eq:bessel_id1}, we simplify this further to
\begin{equation}
\label{eq:J00_STd}
\begin{split}
    J_{lm}^\text{ST} \left( fD, \hat{z} \right)
    = \delta_{m0} \sqrt{\dfrac{2l+1}{4\pi}} \left( 2\pi i^l \int_0^{2\pi fDv} \dfrac{d x}{v} \ e^{i x/v} \left( j_l''(x) + j_l(x) \right) \right) \,.
\end{split}
\end{equation}
The factor we need for the isotropic stochastic gravitational wave background is enclosed in the parenthesis. Making use of the observation $J_{lm}^\text{ST}(fD, \hat{z}) \propto \delta_{m0}$, and performing a three dimensional rotation, we finally get to
\begin{equation}
\label{eq:JlmSTfinal}
\begin{split}
    J^\text{ST}_{lm}\left( fD, \hat{k} \right) =  Y^*_{lm}\left(\hat{k}\right) \left( 2\pi i^l  \int_0^{2\pi fDv} \dfrac{d x}{v} \ e^{i x/v} \left( j_l''(x) + j_l(x) \right) \right) \,.
\end{split}
\end{equation}

Now, moving onto the scalar longitudinal polarization, starting with the contraction \eqref{eq:deps_SL} and performing the sum over $L$ with the same spherical harmonics identities, we end up with
\begin{equation}
\label{eq:J00_SLc}
\begin{split}
    \dfrac{J_{lm}^\text{SL} \left( fD, \hat{z} \right)}{\sqrt{2}}
    = \ & \delta_{m0} \delta_{l0} \dfrac{2\sqrt{ \pi}}{3} \int_0^{2\pi fDv} \dfrac{d x}{v} \ e^{i x/v} \left( \dfrac{j_0(x)}{2} - j_2(x) \right) + \delta_{m0}\delta_{l1} \dfrac{2 \sqrt{3\pi} i}{5}\int_0^{2\pi fDv} \dfrac{d x}{v} \ e^{i x/v} \left( \dfrac{3 j_1(x)}{2} - j_3(x) \right) \\
    & + \delta_{m0} \Theta\left(l - 2\right) 4\pi i^l \sqrt{\dfrac{2l+1}{4\pi}} \int_0^{2\pi fDv} \dfrac{d x}{v} \ e^{i x/v} \left[ \dfrac{d}{dx} \left( \dfrac{j_l(x)}{x} \right) - \dfrac{(l-1)(l+2)}{2}  \dfrac{j_l(x)}{x^2} +  \dfrac{j_l(x)}{2} \right] \,.
\end{split}
\end{equation}
By using the Bessel function differential equation and identity, we are further able to express this simply as
\begin{equation}
\label{eq:J00_SLd}
\begin{split}
    \dfrac{J_{lm}^\text{SL} \left( fD, \hat{z} \right)}{\sqrt{2}}
    = - \delta_{m0} \sqrt{\dfrac{2l+1}{4\pi}} \left( 2\pi i^l \int_0^{2\pi fDv} \dfrac{d x}{v} \ e^{i x/v} j_l''\left(x\right) \right) \,.
\end{split}
\end{equation}
Rotating the $\hat{z}$ axis to a general $\hat{k}$ direction, we get to
\begin{equation}
\label{eq:JlmSLfinal}
\begin{split}
    \dfrac{J^\text{SL}_{lm}\left( fD, \hat{k} \right)}{\sqrt{2}} =  - Y^*_{lm}\left(\hat{k}\right) \left( 2\pi i^l  \int_0^{2\pi fDv} \dfrac{d x}{v} \ e^{i x/v} j_l''\left(x\right) \right) \,.
\end{split}
\end{equation}
We proceed to calculate the isotropic stochastic gravitational wave background's overlap reduction function using the magnitude in the parenthesis. As with the tensor and vector polarizations, we evaluate these integrals numerically to compute the scalar power spectrum.

We note that the scalar transverse integral can be recast as a total boundary for $v = 1$. This can be realized by writing
\begin{equation}
\label{eq:JS_boundary}
    e^{ix/v} \left( j_l''(x) + j_l(x) \right) = \dfrac{d}{dx} \left[ e^{ix/v} \left( j_l'(x) - \dfrac{i}{v} j_l(x) \right) \right] + \dfrac{v^2 - 1}{v^2} e^{ix/v} j_l\left(x\right) \,,
\end{equation}
which reduces to a boundary term if $v = 1$. Therefore by noting the asymptotic expansion
\begin{equation}
    e^{ix/v} \left( j_l'(x) - \dfrac{i}{v} j_l(x) \right) \sim \dfrac{\sqrt{\pi } 2^{-(l+1)}}{\Gamma \left(l+(3/2)\right)} x^l \left( \dfrac{l}{x} + \dfrac{i}{v}(l - 1) + O\left( x \right) \right) \ \ , \ \ \ \ x \rightarrow 0^+
\end{equation}
and the integral identity
\begin{equation}
    \int_0^{r} d x \ e^{i x} j_l\left(x\right) = 2^l r^{l+1} \Gamma (l+1)^2 \, _2\tilde{F}_2(l+1,l+1;l+2,2 l+2;2 i r) \ \ , \ \ \ \ \text{Re}(l) > -1 \,, 
\end{equation}
we may obtain the following analytical expressions for finite $fD$ and $v = 1$:
\begin{equation}
\int_0^{2\pi fD} d x \ e^{i x} \left( j_l''(x) + j_l(x) \right) = e^{2\pi i fD} \left[ j_l'\left(2\pi fD\right) - i j_l\left(2\pi fD\right) \right]
\end{equation}
and
\begin{equation}
\begin{split}
\int_0^{2\pi fD} d x \ e^{i x} j_l''(x) = \, & e^{2\pi i fD} \left[ j_l'\left(2\pi fD\right) - i j_l\left(2\pi fD\right) \right] \\
& \ \ - 2^l \left(2\pi fD\right)^{l+1} \Gamma (l+1)^2 \, _2\tilde{F}_2 \left( l+1,l+1;l+2,2 l+2;4\pi i fD \right) \,,
\end{split}
\end{equation}
where $\, _2\tilde{F}_2 \left( a, b; c, d; x \right) = \, _2{F}_2 \left( a, b; c, d; x \right)/\left( \Gamma(c) \Gamma(d) \right)$ is a regularized hypergeometric function. These help significantly to reduce the numerical evaluation time of the power spectra, at least for $v = 1$. Analytical expressions for arbitrary $v$ in the infinite distance limit may also be obtained by utilizing \eqref{eq:JS_boundary} and noting that
\begin{equation}
    \int_0^\infty \dfrac{dx}{v} \ e^{ix/v} j_l(x) = \sqrt{\pi } 2^{-(l+1)} (i v)^{l+1} \Gamma (l+1) \, _2\tilde{F}_1\left(\frac{l+1}{2},\frac{l+2}{2};l+\frac{3}{2};v^2\right) \,.
\end{equation}
This may be used to speed up the numerical integration for the infinite distance limit.

\subsection{ORF and power spectra}
\label{subsec:power_spectra_scalar}

We take the calculated magnitudes in the previous section to compute the overlap reduction function for an isotropic stochastic gravitational wave background \eqref{eq:orf_general}. This way, doing it separately for the scalar transverse and scalar longitudinal polarizations, using the addition theorem \eqref{eq:addition_theorem}, we get to the result
\begin{equation}
    \gamma_{ab} \left( \zeta, fD_i \right) = \sum_l \dfrac{2l+1}{4\pi} C_l P_l\left( \cos \zeta \right) \,,
\end{equation}
where the scalar power spectrum multipoles $C_l$ are given by
\begin{equation}
    C_l = \dfrac{32 \pi^2 F_l\left(f D_a \right) F_l^*\left( fD_b\right)}{\sqrt{4\pi}} \, 
\end{equation}
with the quantity $F_l(fD)$ being
\begin{equation}
    F_l(fD) = - \dfrac{i}{2} \int_0^{2\pi fDv} \dfrac{dx}{v} \ e^{ix/v} R_l\left(x\right) \,.
\end{equation}
In the above expression, $R_l^\text{SL}(x) = j_l''(x)$ for the scalar longitudinal polarization and $R_l^\text{ST}(x) = -\left( R_l^\text{SL}(x) + j_l(x) \right)/\sqrt{2}$ for the scalar transverse polarization. These $R_l(x)$ functions can be confirmed to be same ones singled out in \cite{Qin:2020hfy} in their Appendix A. The quantity $F_l(\infty)$ are the projection factors considered in \cite{Qin:2020hfy} such that $C_l \propto 32 \pi^2 F_l(\infty) F_l^*(\infty)$. We highlight the main difference to be that the upper limit of the integral is finite which keeps the power spectra defined for either polarizations.

In the limit $fD \rightarrow \infty$ and $v \rightarrow 1$, for the scalar transverse monopole and dipole, it can be checked that $F_0^\text{ST}(\infty) = -1/\left(2\sqrt{2}\right)$ and $F_1^\text{ST}(\infty) = -i/\left(6\sqrt{2}\right)$. In the same limit, the scalar longitudinal monopole and dipole projection factors become undefined. On the other hand, in the infinite pulsar distance limit, $fD \rightarrow \infty$, but with arbitrary group velocity $v$, the higher order multipoles $l \geq 2$ become constrained as $F^\text{ST}_l\left(\infty\right) + \left( 1 - v^2 \right) \left( F^\text{SL}_l\left(\infty\right)/\sqrt{2} \right) = 0$. All these agree with \cite{Qin:2020hfy}.

We calculate the autocorrelation using the power spectrum and the real space formalism to assess the validity of our power spectrum calculation. The antenna pattern functions for the scalar transverse and longitudinal modes are
\begin{equation}
    F^\text{ST}_a \left( \hat{k} = (\theta, \phi) \right) = \dfrac{\sin^2 \theta}{2 \left( 1 + v \cos \theta \right)}
\end{equation}
and
\begin{equation}
    F^\text{SL}_a \left( \hat{k} = (\theta, \phi) \right) = \dfrac{\cos^2\theta}{\sqrt{2} \left( 1 + v \cos \theta \right)} \,.
\end{equation}
The scalar autocorrelation reduces to the integrals
\begin{equation}
\label{eq:gammaaaST}
    \gamma_{aa}^\text{ST} = \int_0^\pi \dfrac{d\theta}{\sqrt{4\pi}} \left( \frac{2 \pi  \sin ^5 \theta \sin ^2 ( \pi fD (1 + v \cos \theta ))}{(1 + v \cos \theta)^2} \right) 
\end{equation}
and
\begin{equation}
\label{eq:gammaaaSL}
    \gamma_{aa}^\text{SL} = \int_0^\pi \dfrac{d\theta}{\sqrt{4\pi}} \left( \frac{4 \pi  \sin \theta \cos ^4 \theta \sin ^2 \left(\pi fD (1 + v \cos \theta ) \right)}{(1 + v \cos \theta )^2} \right) \,. 
\end{equation}
It is interesting that $\gamma_{aa}^\text{ST}$ coincides with the transverse tensor $\gamma_{aa}^T$.

\subsection{Phenomenology}
\label{subsec:scalar_phenomenology}

\begin{figure}[h!]
\center
\begin{adjustbox}{minipage = \linewidth, scale = 0.95}
	\subfigure[ \, scalar transverse, $v = 0.99$, $C_2^{fD=100} = 0.0068, C_2^{fD=500} = 0.0069, C_2^{fD=\infty} = 0.0069$ ]{
		\includegraphics[width = 0.45 \textwidth]{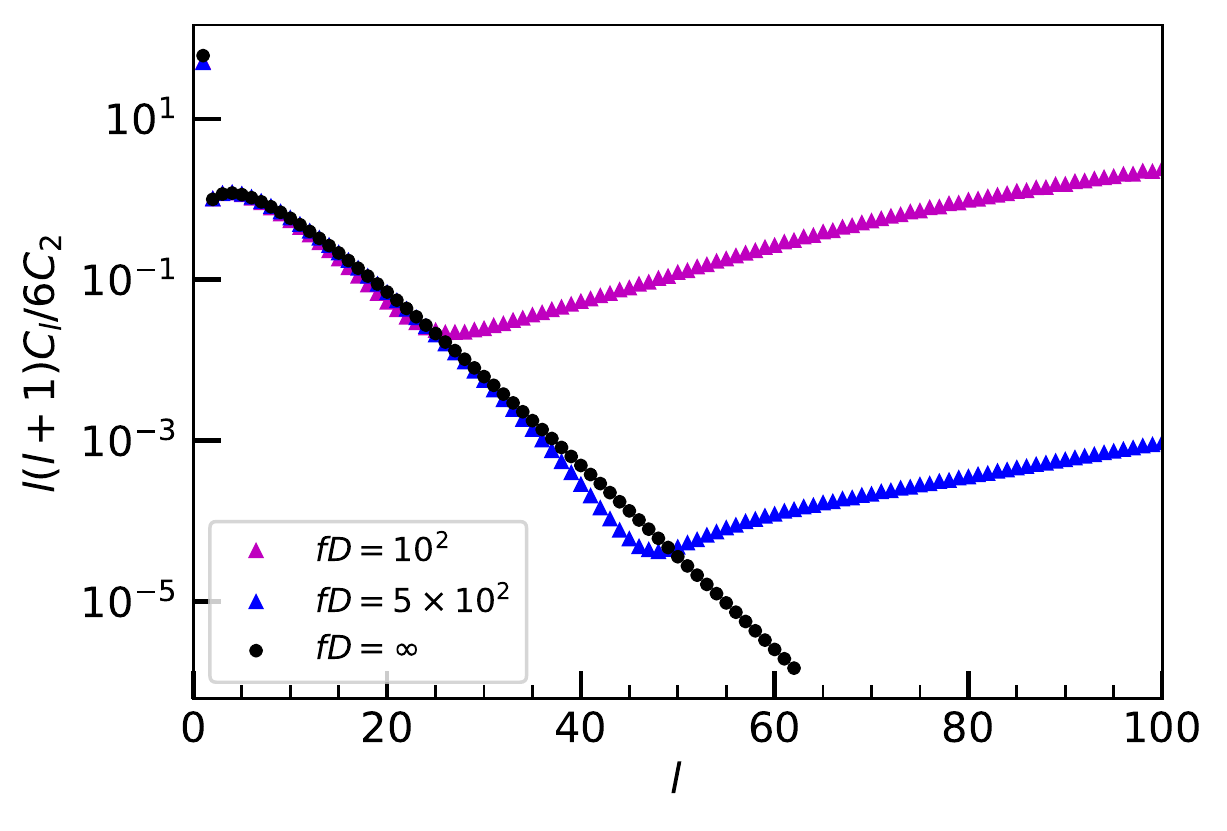}
		}
	\subfigure[ \, scalar transverse, $v = 0.99$, $\gamma_{aa}^{fD=100} = 2.17, \gamma_{aa}^{fD=500} = 2.17, \gamma_{aa}^{fD=\infty} = 2.17$ ]{
		\includegraphics[width = 0.45 \textwidth]{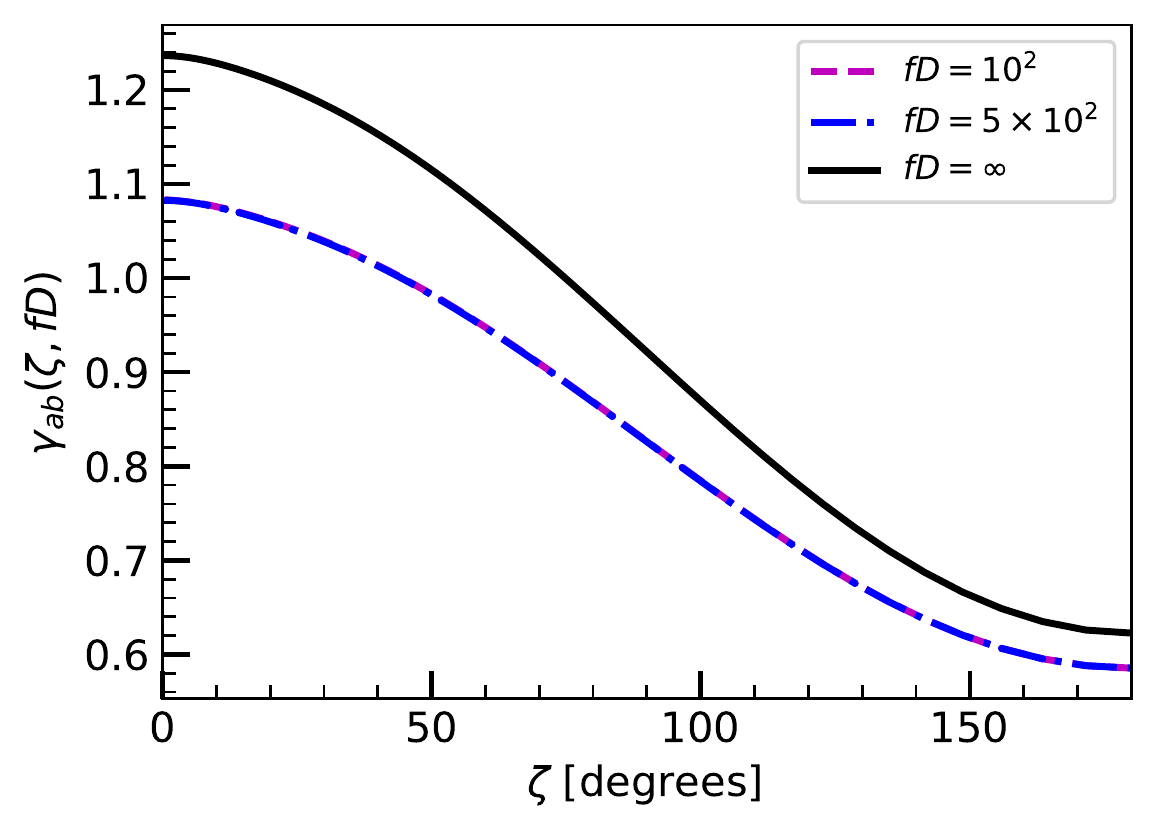}
		}
	\subfigure[ \, scalar transverse, $v = 1/2$, $C_2^{fD=100} = 0.18, C_2^{fD=500} = 0.18, C_2^{fD=\infty} = 0.18$ ]{
		\includegraphics[width = 0.45 \textwidth]{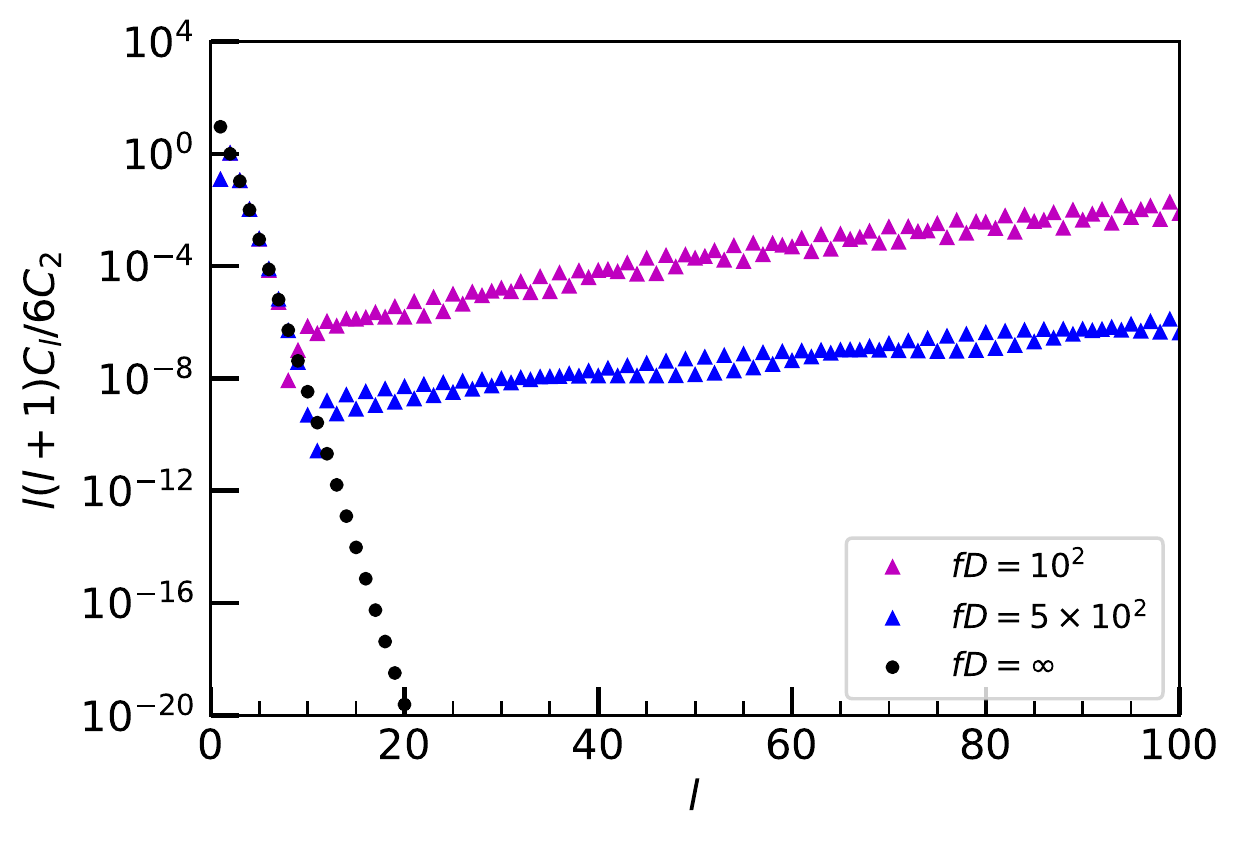}
		}
	\subfigure[ \, scalar transverse, $v = 1/2$, $\gamma_{aa}^{fD=100} = 1.06, \gamma_{aa}^{fD=500} = 1.06, \gamma_{aa}^{fD=\infty} = 1.06$ ]{
		\includegraphics[width = 0.45 \textwidth]{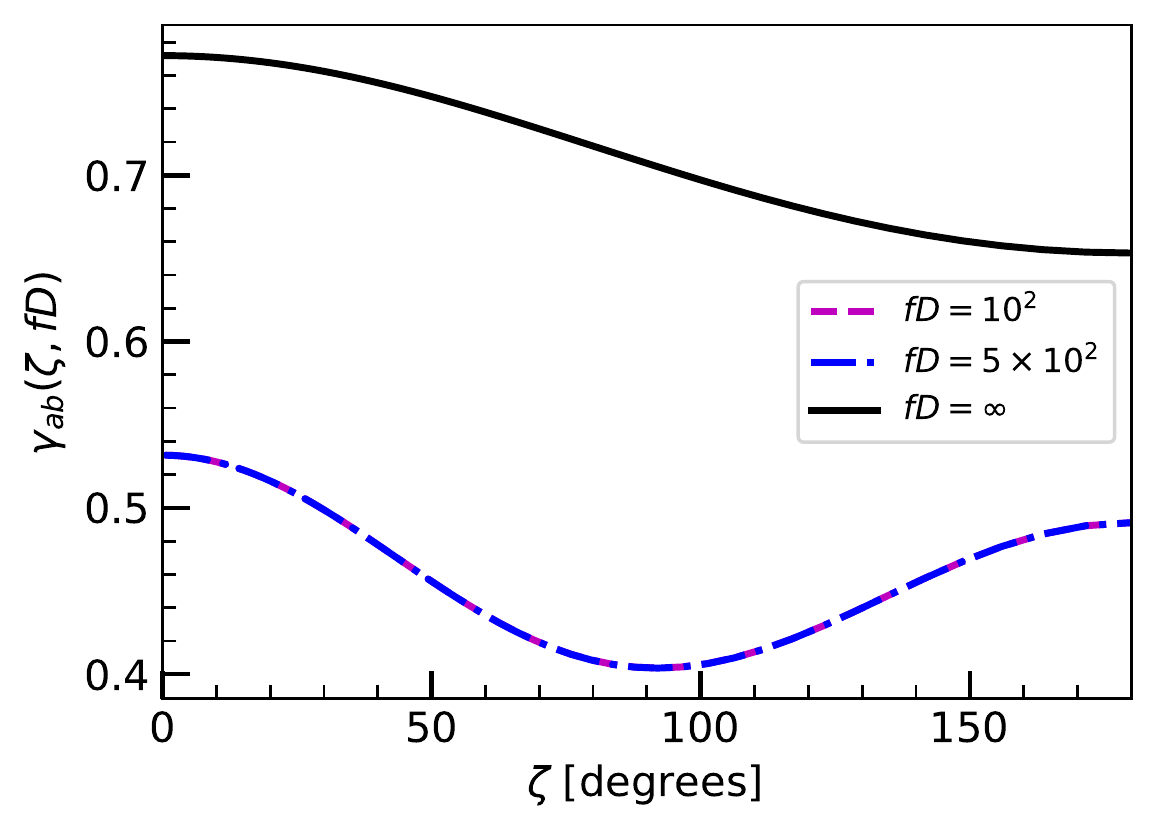}
		}
	\subfigure[ \, scalar transverse, $v = 1/100$, $C_2^{fD=100} = 0.12, C_2^{fD=500} = 0.19, C_2^{fD=\infty} = 0.20$ ]{
		\includegraphics[width = 0.45 \textwidth]{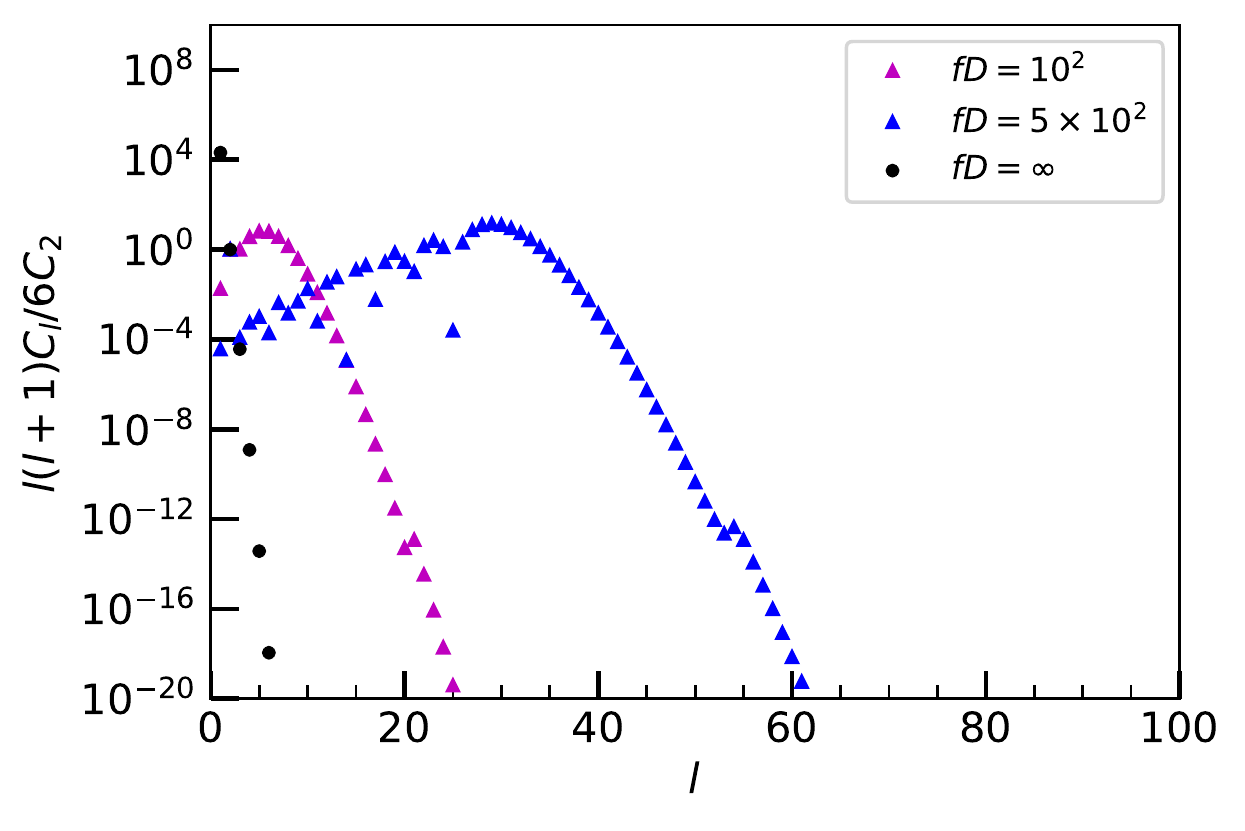}
		}
	\subfigure[ \, scalar transverse, $v = 1/100$, $\gamma_{aa}^{fD=100} = 0.97, \gamma_{aa}^{fD=500} = 0.95, \gamma_{aa}^{fD=\infty} = 0.95$ ]{
		\includegraphics[width = 0.45 \textwidth]{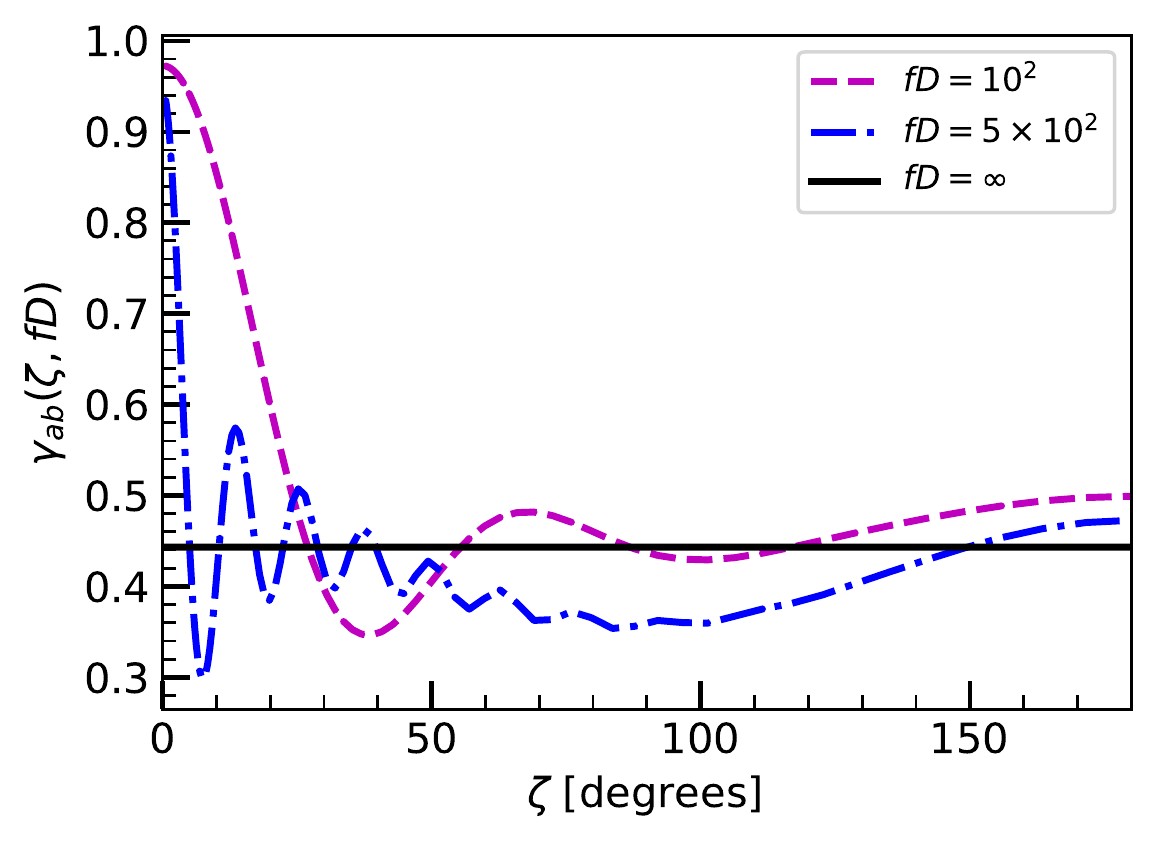}
		}
\caption{The isotropic power spectra multipoles $C_l$ and the overlap reduction functions $\gamma_{ab}\left(\zeta, fD \right)$ of the scalar transverse polarization with group velocities $v = 99/100$, $v = 1/2$, and $v = 1/100$. The overlap reduction functions were constructed with only the first {sixty} multipoles. The $fD = \infty$ curves in (d) and (f) were divided by $20$ and $2 \times 10^8$, respectively. The autocorrelation $\gamma_{aa}$'s are computed from \eqref{eq:gammaaaST}.
}
\label{fig:ClST}
\end{adjustbox}
\end{figure}

We present the power spectra and resulting overlap reduction functions individually for each of the scalar polarizations. Figure \ref{fig:ClST} shows this for the scalar transverse polarization with $v \sim 1$, $v = 1/2$, and $v = 10^{-2}$ at various pulsar distances.

This again echoes the important physical difference between the finite and infinite pulsar distance cases. In the infinite case \cite{Qin:2018yhy, Qin:2020hfy}, the power spectra drops continuously at large $l$, or small angles, as $C_l \sim 1/l^k$ for some positive $k$. This implies that the correlations vanish for pulsars that are infinitesimally separated in the sky. However, in a real setting, pulsars are separated at a finite line of sight distance from the observer (`Us'). Figure \ref{fig:ClST}(a) shows that at some $l \sim 20 - 50$ with a nearly luminal scalar degree of freedom ($v \sim 1$), the power spectra ceases to drop and instead sustains a slow increase. This tells that correlations at small angular separations are strengthened for nearby pulsars, which makes sense for neighboring astrophysical sources \cite{Bernardo:2022vlj}. We recognize this partial sustenance in the transverse-traceless tensor as well as the vector polarizations for finite pulsar distances in the previous sections. The corresponding {ORF} is shown in Figure \ref{fig:ClST}(b). This presents the scalar transverse signal resembling what looks like a dipole \cite{NANOGrav:2020bcs} understandably because its power spectrum is dominated by the dipole. In this near luminal limit, we also find that the finite pulsar distance curves by themselves are not so much distinguishable, as in the tensor and vector cases. However, the finite and infinite pulsar distance cases are visually distinguishable, unlike their tensor and vector counterparts.

Figures \ref{fig:ClST}(c-d) show the multipoles and the {ORFs} when the scalar modes propagate at half the speed of light. In this case, we find that the power spectra multipoles feature an overall decrease in magnitude, and comes with an even sharper drop at small $l$. This is reflected in the {ORF} which tends to flatter values, obviously being dominated by the monopole, as compared with the near luminal scalar case. Take note that the {ORFs} for large angles relevant for pulsar timing array for the finite pulsar distance cases are visually indistinguishable as displayed by their multipoles (Figures \ref{fig:ClST}(c)). It is worth noting that the monopole in this velocity in the infinite distance limit diverges. In Figure \ref{fig:ClST}(d), the infinite distance curve was as a matter of fact divided by twenty in order to be shown together with the other signals. This divergence will be even more drastic at lower velocities.

When the velocity is decreased further to the nonrelativistic limit and infinite distance limit, the power spectrum multipoles drop sharper beginning with the monopole, and so the {ORF} reduces to practically a flat horizontal line. However, for the finite distance cases, at some point, we find that the dipole, and even the quadrupole, becomes suppressed compared to the succeeding low multipoles until a peak of the power spectrum appears. This manifests itself as an oscillation in the {ORF}, at an angle $\zeta = 180^\circ/l_\text{peak}$ defined by the multipole number $l_\text{peak}$ at which the power spectrum peaks. Figures \ref{fig:ClST}(e-f) explicitly show this with $v = 0.01$. As alluded, for this case, the low multipoles for the finite distance cases can now be distinguished, as contributions beyond the monopole and dipole become significant, and this manifests in the {ORF} at large angular separations. A drastic technical difference in the {ORF} between the finite and infinite pulsar distance case with $v \ll 1$ can also be realized. The {ORF} for the infinite distance limit in Figure \ref{fig:ClST}(f) was divided by $2 \times 10^8$ to be comparable with the other curves. Further, whereas the infinite distance limit signal reduces to a monopole (a mere horizontal line), the finite distance cases present oscillations owing to the fact that pulsars are astrophysical objects in an observable universe.

\begin{figure}[h!]
\center
\begin{adjustbox}{minipage = \linewidth, scale = 0.95}
	\subfigure[ \, scalar longitudinal, $v = 99/100$, $C_2^{fD=100} = 34.2, C_2^{fD=500} = 34.7, C_2^{fD=\infty} = 34.8$ ]{
		\includegraphics[width = 0.45 \textwidth]{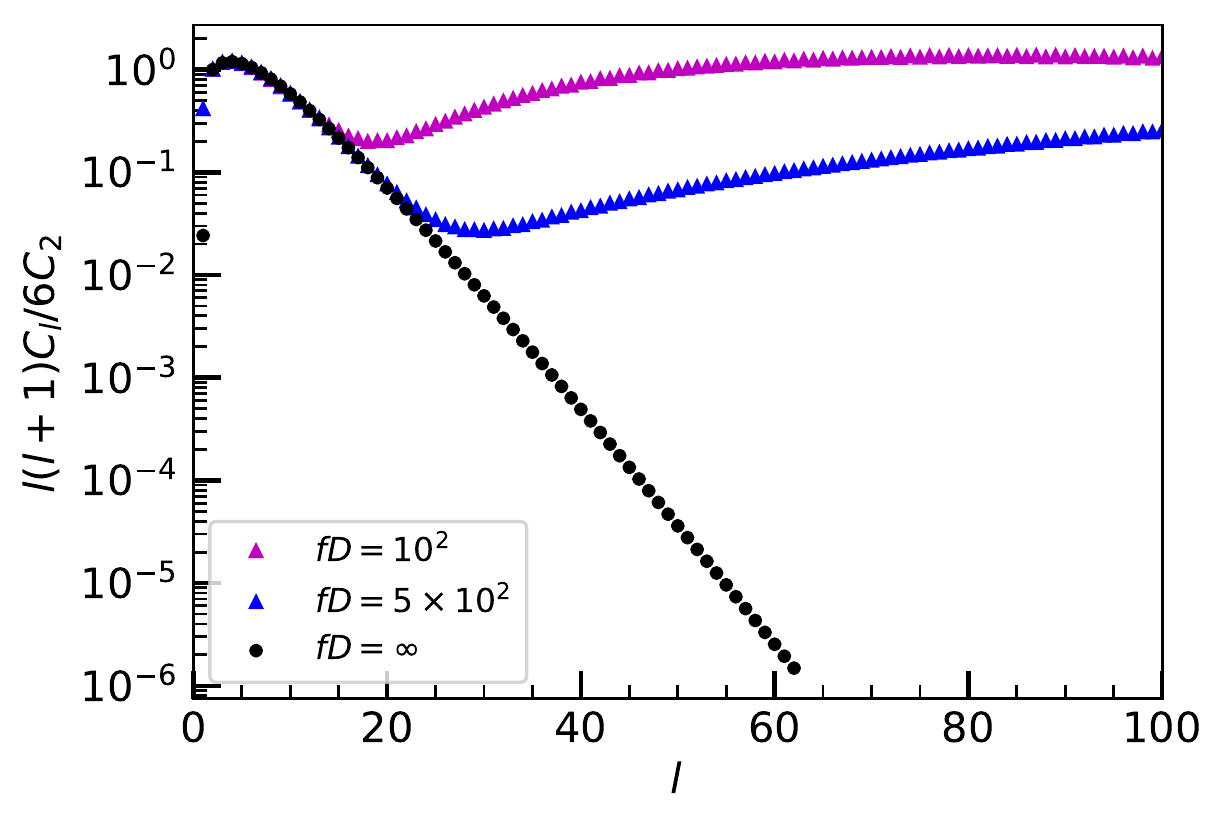}
		}
	\subfigure[ \, scalar longitudinal, $v = 99/100$, $\gamma_{aa}^{fD=100} = 151, \gamma_{aa}^{fD=500} = 158, \gamma_{aa}^{fD=\infty} = 158$ ]{
		\includegraphics[width = 0.45 \textwidth]{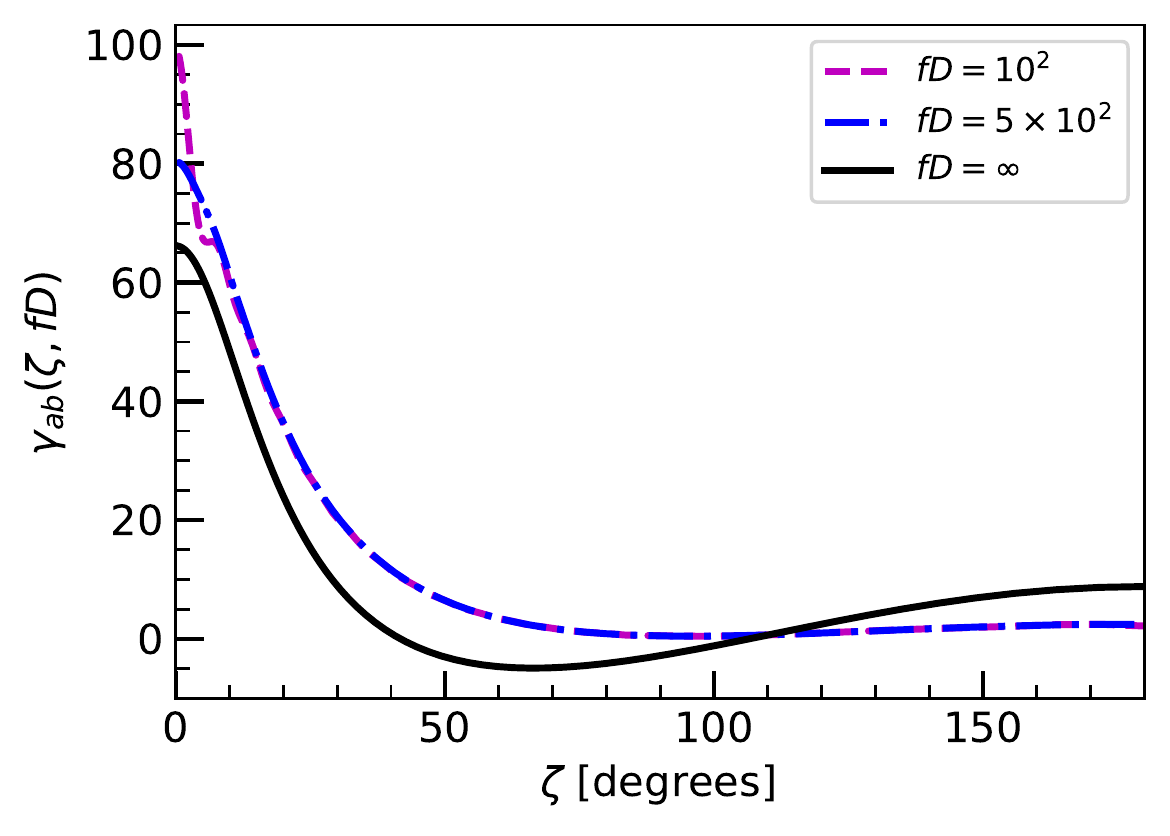}
		}
	\subfigure[ \, scalar longitudinal, $v = 1/2$, $C_2^{fD=100} = 0.64, C_2^{fD=500} = 0.64, C_2^{fD=\infty} = 0.64$ ]{
		\includegraphics[width = 0.45 \textwidth]{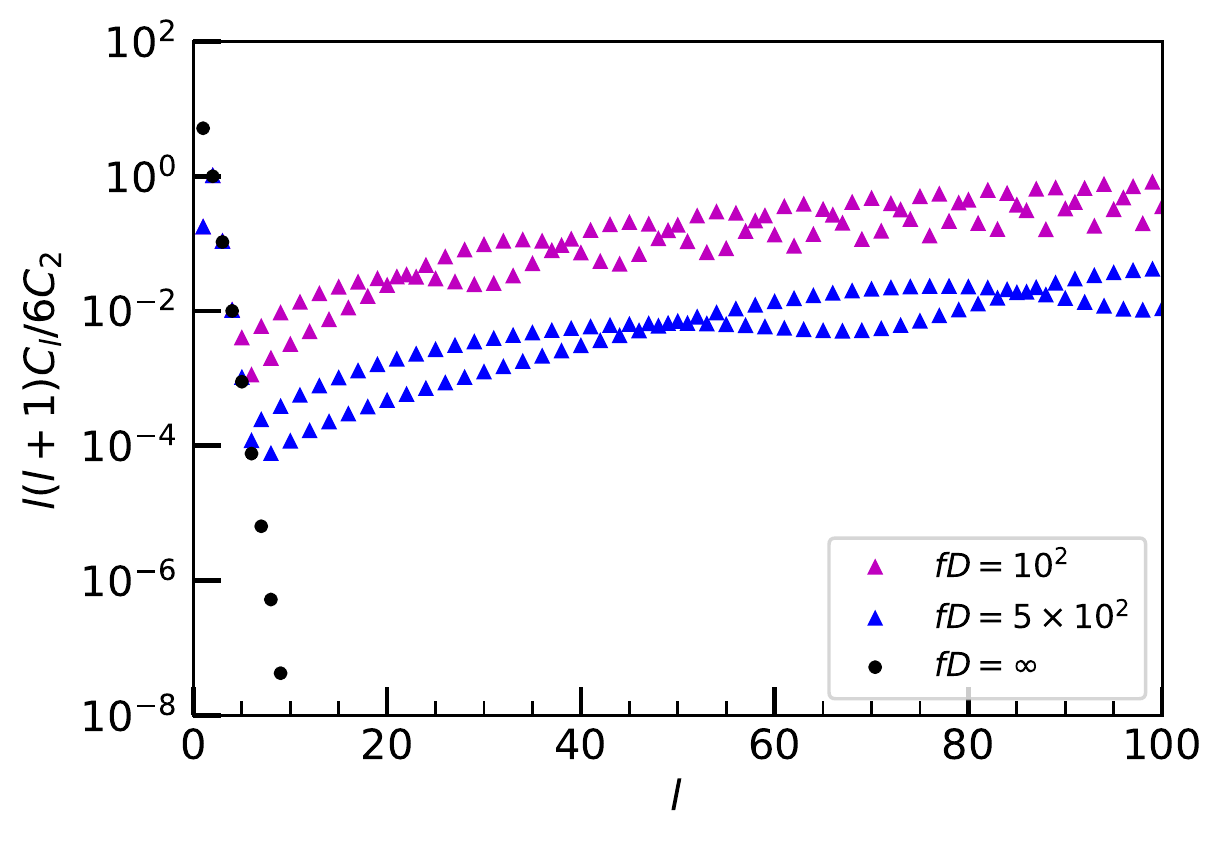}
		}
	\subfigure[ \, scalar longitudinal, $v = 1/2$, $\gamma_{aa}^{fD=100} = 1.26, \gamma_{aa}^{fD=500} = 1.26, \gamma_{aa}^{fD=\infty} = 1.26$ ]{
		\includegraphics[width = 0.45 \textwidth]{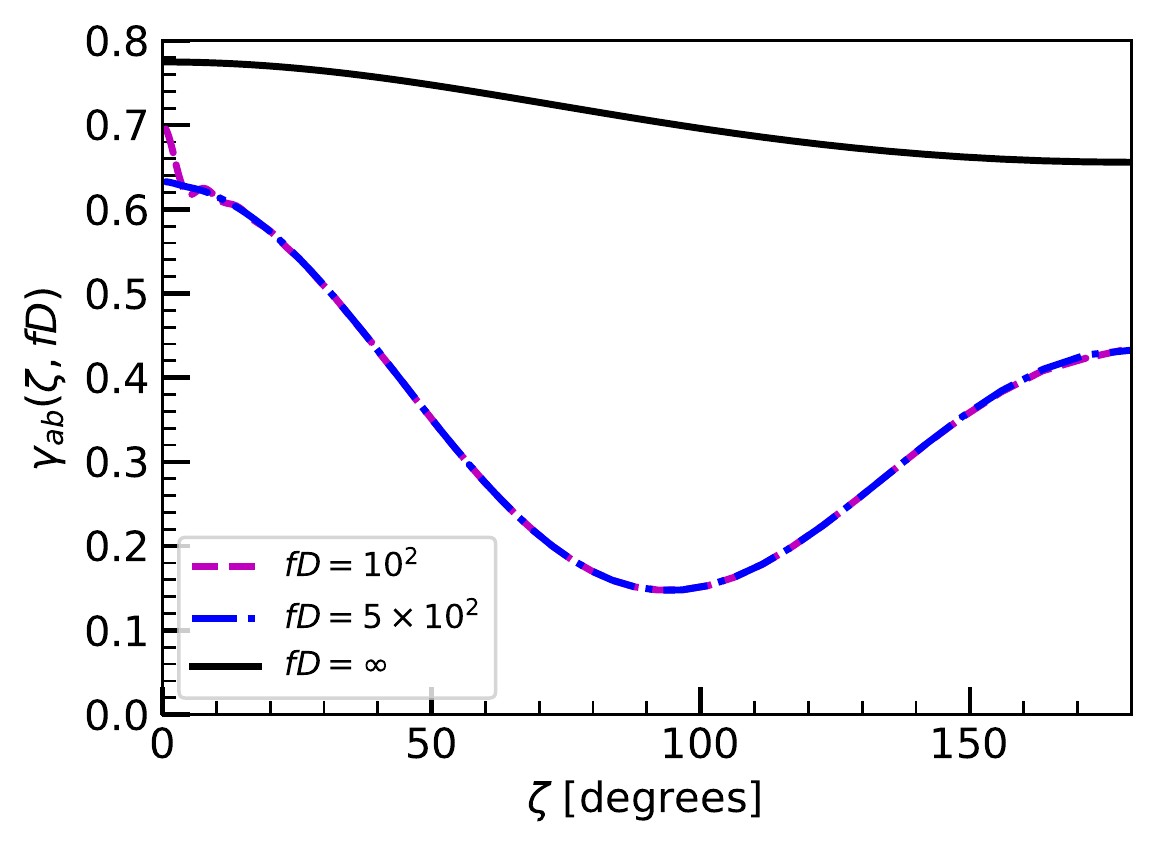}
		}
	\subfigure[ \, scalar longitudinal, $v = 1/100$, $C_2^{fD=100} = 0.020, C_2^{fD=500} = 0.367, C_2^{fD=\infty} = 0.396$ ]{
		\includegraphics[width = 0.45 \textwidth]{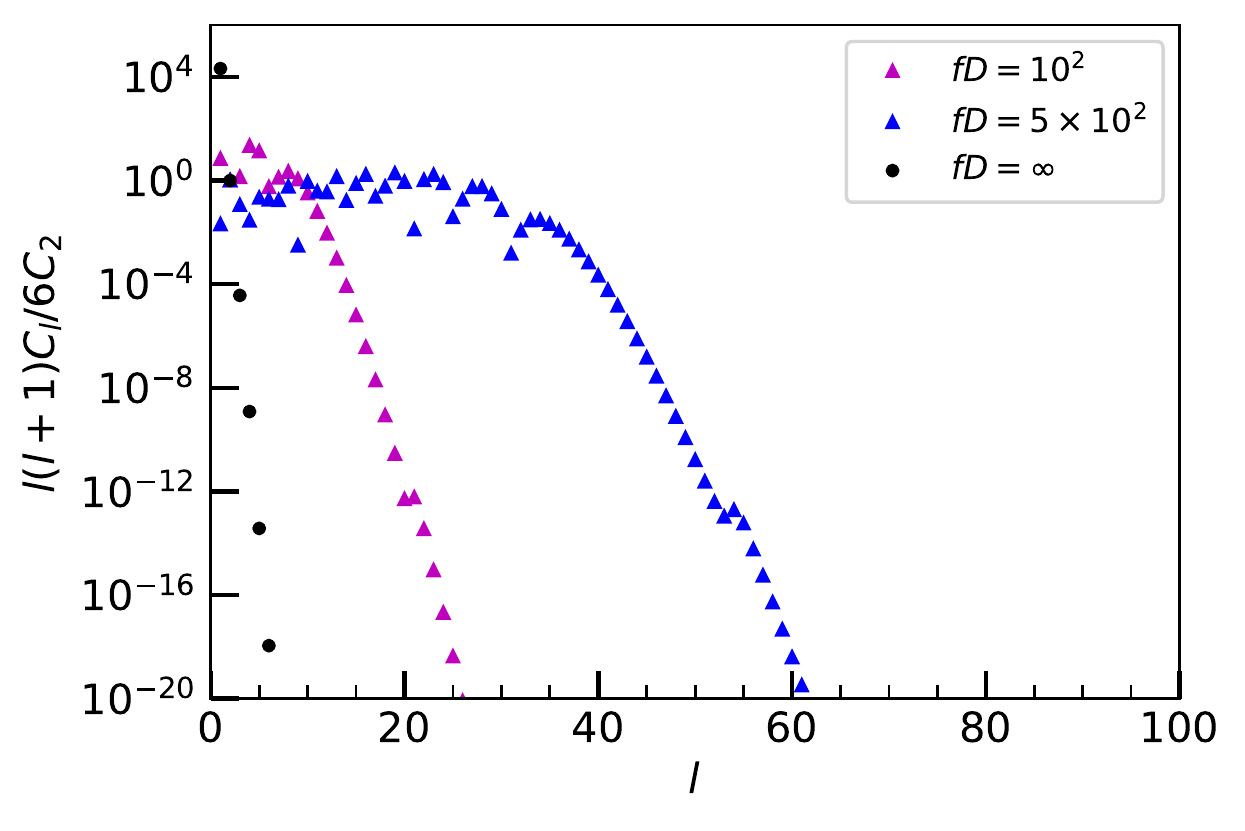}
		}
	\subfigure[ \, scalar longitudinal, $v = 1/100$, $\gamma_{aa}^{fD=100} = 0.40, \gamma_{aa}^{fD=500} = 0.69, \gamma_{aa}^{fD=\infty} = 0.71$ ]{
		\includegraphics[width = 0.45 \textwidth]{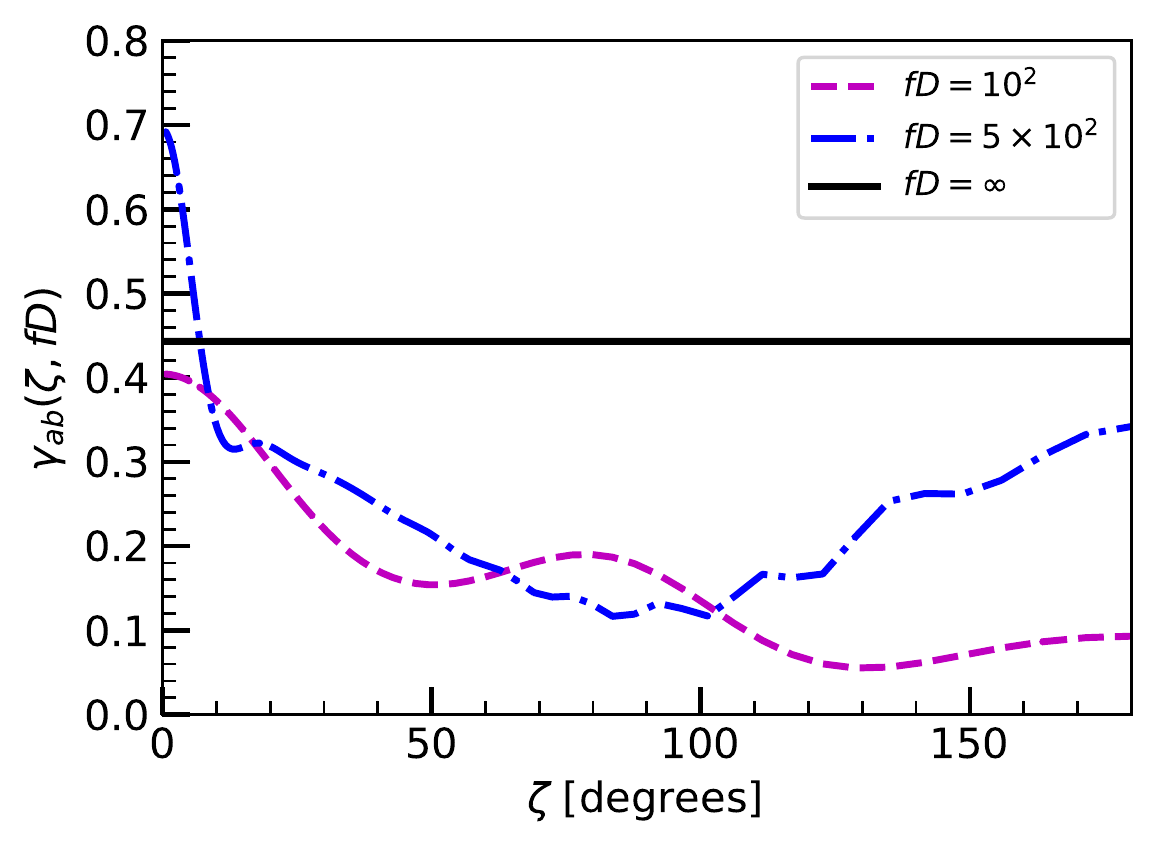}
		}
\caption{The isotropic power spectra multipoles $C_l$ and the overlap reduction functions $\gamma_{ab}\left(\zeta, fD \right)$ of the scalar longitudinal polarization with group velocities $v \sim 1$, $v = 1/2$, and $v = 1/100$. The overlap reduction functions were constructed with only the first {sixty} multipoles. The $fD = \infty$ curves in (d) and (f) were divided by $40$ and $4 \times 10^8$, respectively. The autocorrelation $\gamma_{aa}$'s were calculated from \eqref{eq:gammaaaSL}.
}
\label{fig:ClSL}
\end{adjustbox}
\end{figure}

Figure \ref{fig:ClSL} shows the power spectra multipoles and the {ORFs} for the scalar longitudinal polarization with the previous choices for the velocity.

First off, we recall that the scalar longitudinal mode is undefined with $v = 1$ and infinite pulsar distances. For this alone, we realize the practical advantage of keeping the pulsars at finite distances. In \cite{Bernardo:2022vlj}, we even find compelling statistical evidence of scalar longitudinal polarization in pulsar timing array data.

In the nearly luminal scalar case (Figures \ref{fig:ClSL}(a-b)), we find the power spectrum to be dominated by the low multipoles aside from the dipole and the quadrupole. This translates to the particular shape of the {ORF} at low angles, as shown. As with the scalar transverse polarization, the power drop is eventually disrupted at some sufficiently large $l$, or small angles, where real, neighboring pulsars may be correlated in their history. However, in the infinite pulsar distance case, the dipole drops compared to the monopole and others close by, resulting in the overlap reduction function being distinguishable for large angles compared to the finite pulsar distance cases. This large angular separation distinction between the finite and infinite pulsar distance cases also manifests at other velocities. We find the similar behavior in the scalar transverse case where the monopolar and dipolar powers numerically diverge as seen in the power spectrum plots.

Again, as in the scalar transverse polarization, at half the speed of light, the scalar longitudinal power spectra feature a steeper drop beginning with the quadrupole as $l$ increases (Figures \ref{fig:ClSL}(c-d)). However, it can be seen that in the infinite distance case, the dipolar contribution is larger than the quadrupole, while for the finite distance cases, the dipolar power is otherwise suppressed. This clearly reflects in the {ORFs} where the infinite distance curve can be seen to be shaped like a dipole, while the finite distance curves look like a quadrupole. We may mention that the monopole also starts to contribute significantly in the infinite distance case at subluminal velocities \cite{Qin:2020hfy}. The infinite distance curve in Figure \ref{fig:ClSL}(d) was divided by forty to be comparable with the other curves. In this case, the infinite distance correlation becomes mainly dominated by the monopole and the dipole, and so the corresponding {ORF} shapes appear like the dipole at half the speed of light. The finite pulsar distance cases, being more dominantly sourced by the quadrupole, instead show a significant departure from the infinite distance limit, as now their shape resembles the Hellings-Downs curve. This time, also, the {ORFs} for the finite pulsar distance cases very much coincide.

We find that the situation changes in the nonrelativistic scalar ($v = 1/100$), in Figures \ref{fig:ClSL}(e-f). In this case, in the infinite distance limit, the power spectrum is significantly dominated by the monopole, which is reflected in the {ORF} being practically a horizontal line \cite{Qin:2020hfy}. As we have also alluded a while ago, the monopole numerically diverges at subluminal velocities, such that we had to divide the infinite distance curve in Figure \ref{fig:ClSL}(f) by a factor $O\left(10^8\right)$. The situation also manifested in the scalar transverse case. On the other hand, for finite pulsar distances, all computations remain well at hand. In this case, we find that most of the low multipoles continue to contribute to the power spectrum, giving a nontrivial angular correlation, and so producing an {ORF} that is distinguishable compared with the infinite distance limit. It is notable that the {ORFs} for the finite distance cases are also distinguishable between themselves. This is teased by the scalar power spectrum, which drops significantly beginning at $l \sim 10$ for $fD = 100$ and beginning at $l \sim 30$ for $fD = 500$. The {ORF} of the scalar longitudinal polarization remains completely nontrivial in all cases for finite distance and subluminal velocities.

\section{Discussion}
\label{sec:discussion}

We have presented a power spectrum method for calculating the overlap reduction function in a pulsar timing array. We argue that this is fast and efficient, particularly with the present data set in which the pulsar pairs have about at least {three} degrees angular separation, thus requiring only the first few of the power spectrum multipoles. This is its main advantage over the real space formalism which is challenged numerically by integration over a pole of the integrand. The two methods lead to the same result regardless, as captured by the autocorrelation computations in Table \ref{tab:GaafD100}.

\begin{table}[h!]
    \centering
    \caption{Autocorrelation $\gamma_{aa}$ calculated using the power spectrum ($l \leq l_\text{max}$) and the real space formalism (RSF) with $fD = 100$. The modes T, V, ST, and SL stand for `tensor', `vector', `scalar transverse', and `scalar longitudinal', respectively. {$t_{\rm eval}$ are the corresponding numerical evaluation times (in seconds) in a 12th Gen Intel Core i7-12700, 2100 Mhz Processor Computer.}}
    \begin{tabular}{|c|c|cc|cc|cc|} \hline
    mode & $\phantom{\dfrac{1}{1}}$ $v$ $\phantom{\dfrac{1}{1}}$ & $\phantom{gg}$ $\gamma_{aa}^{l \leq 30}$ $\phantom{gg}$ & $t_{\rm eval}$ (s) & $\phantom{gg}$ $\gamma_{aa}^{l \leq 1000}$ $\phantom{gg}$ & $t_{\rm eval}$ (s) & $\phantom{gg}$ $\gamma_{aa}^{\text{RSF}}$ $\phantom{gg}$ & $t_{\rm eval}$ (s) \\ \hline \hline
    \multirow{3}*{\phantom{ggg} T \phantom{ggg}} & $\phantom{\dfrac{1}{1}}$ $0.99$ $\phantom{\dfrac{1}{1}}$ & $1.08$ & 0.26 & $2.17$ & 46 &  $2.17$ & 0.03 \\ \cline{2-8}
    & $\phantom{\dfrac{1}{1}}$ $0.50$ $\phantom{\dfrac{1}{1}}$ & $0.53$ & 0.81 & $1.06$ & 11 & $1.06$ & 0.01 \\ \cline{2-8}
    & $\phantom{\dfrac{1}{1}}$ $0.01$ $\phantom{\dfrac{1}{1}}$ & $0.97$ & 0.30 & $0.97$ & 1.0 & $0.97$ & $10^{-3}$ \\ \hline
    \multirow{3}*{\phantom{ggg} V \phantom{ggg}} & $\phantom{\dfrac{1}{1}}$ $0.99$ $\phantom{\dfrac{1}{1}}$ & $7.92$ & 1.4 & $15.7$ & 15 & $15.7$ & 0.03 \\ \cline{2-8}
    & $\phantom{\dfrac{1}{1}}$ $0.50$ $\phantom{\dfrac{1}{1}}$ & $0.67$ & 0.80 & $1.34$ & 6.5 & $1.34$ & 0.01 \\ \cline{2-8}
    & $\phantom{\dfrac{1}{1}}$ $0.01$ $\phantom{\dfrac{1}{1}}$ & $1.20$ & 0.33 & $1.20$ & 0.83 & $1.20$ & $10^{-3}$ \\ \hline
    \multirow{3}*{\phantom{ggg} ST \phantom{ggg}} & $\phantom{\dfrac{1}{1}}$ $0.99$ $\phantom{\dfrac{1}{1}}$ & $1.08$ & 3.9 & $2.17$ & 59 & $2.17$ & 0.02 \\ \cline{2-8}
    & $\phantom{\dfrac{1}{1}}$ $0.50$ $\phantom{\dfrac{1}{1}}$ & $0.53$ & 2.2 & $1.06$ & 20 & $1.06$ & 0.01 \\ \cline{2-8}
    & $\phantom{\dfrac{1}{1}}$ $0.01$ $\phantom{\dfrac{1}{1}}$ & $0.97$ & 1.1 & $0.97$ & 2.1 & $0.97$ & $10^{-3}$ \\ \hline
    \multirow{3}*{\phantom{ggg} SL \phantom{ggg}} & $\phantom{\dfrac{1}{1}}$ $0.99$ $\phantom{\dfrac{1}{1}}$ & $81.8$ & 3.3 & $151$ & 45 & $151$ & 0.03 \\ \cline{2-8}
    & $\phantom{\dfrac{1}{1}}$ $0.50$ $\phantom{\dfrac{1}{1}}$ & $0.65$ & 2.3 & $1.26$ & 15 & $1.26$ & 0.01 \\ \cline{2-8}
    & $\phantom{\dfrac{1}{1}}$ $0.01$ $\phantom{\dfrac{1}{1}}$ & $0.40$ & 0.80 & $0.40$ & 1.6 & $0.40$ & $10^{-3}$ \\ \hline
    \end{tabular}
    \label{tab:GaafD100}
\end{table}

The autocorrelation can be interpreted as the correlation of a pulsar with itself and embodies the small scale power in a pulsar timing array. To compute this using the power spectrum formalism requires at least a few thousand multipoles to guarantee a degree of numerical convergence. This is shown in Table \ref{tab:GaafD100} for $fD = 100$, or $D \sim 30$ parsecs, where it can be seen that the large scale multipoles ($l \leq 30$ or $\zeta \geq 6^\circ$), those which we have utilized in the overlap reduction functions in the previous sections, only capture nearly half of the total power in small scales ($ l \leq 1000$ or $\zeta \geq 0.18^\circ$). We confirm that the real space formalism gets to the same numbers. Understandably this is because the real space formalism involves both earth and pulsar terms, thereby incorporating the small and large scale power at the same time. The situation deviates in the nonrelativistic modes ($v \ll 1$) where nearly the full power is captured by the low multiples ($l \lesssim 100$), but this can be explained quite easily by looking at the power spectrum. The power peaks and drops sharply within the first hundred multipoles for nonrelativistic polarization modes ($v \ll 1$). Figures \ref{fig:ClT}, \ref{fig:ClV}, \ref{fig:ClST}, and \ref{fig:ClSL} support this point that the low multipoles by themselves effectively capture the full power at nonrelativistic speeds.

\begin{figure}[h!]
    \centering
    \includegraphics[width = 0.45 \textwidth]{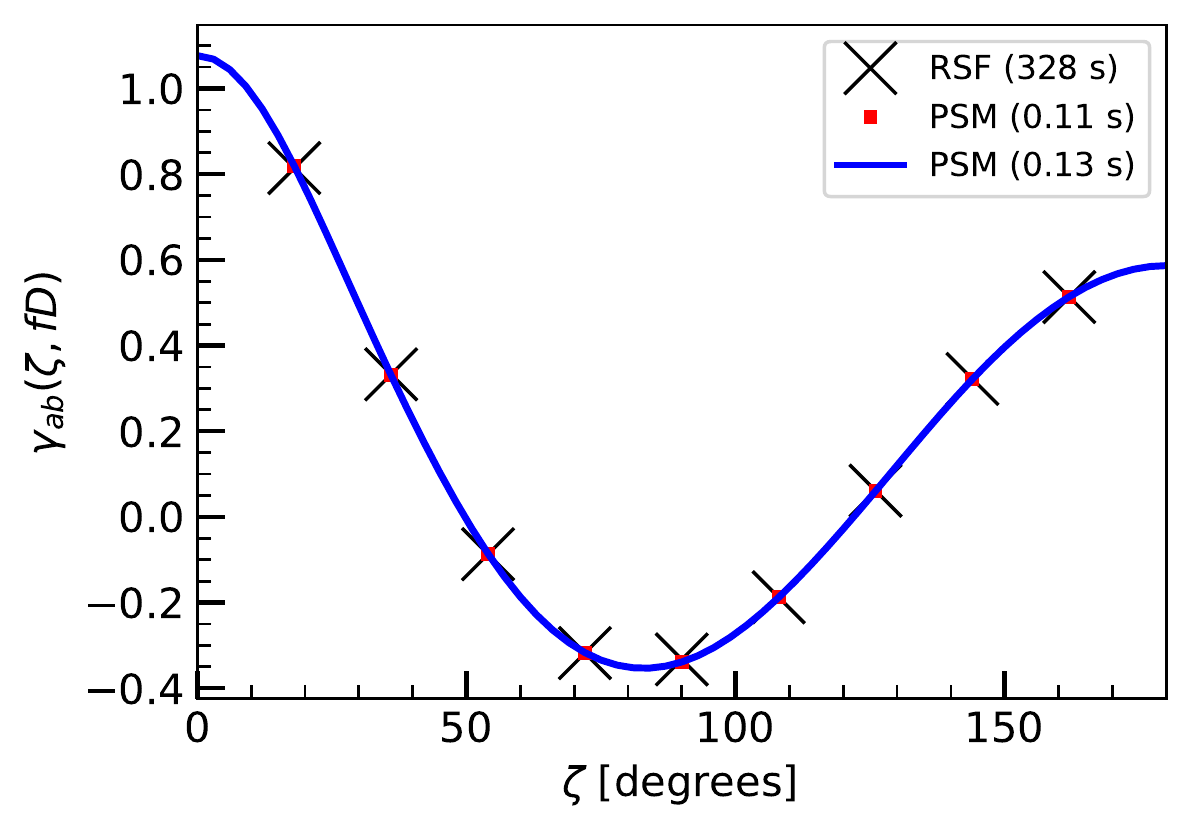}
    \caption{{The overlap reduction function produced by near luminal tensor polarizations ($v = 0.99$) computed using the real space formalism (black crosses) and the power spectrum method (red squares) at eight equally spaced points in $\zeta \in (0^\circ, 180^\circ)$. The blue solid curve is evaluated at a grid with sixty points. Legends show in the parenthesis the corresponding total evaluation times in a 12th Gen Intel Core i7-12700, 2100 Mhz Processor Computer.}}
    \label{fig:orfTv1_PSvsRSF}
\end{figure}

{We also highlight in Table \ref{tab:GaafD100} the advantage of the real space formalism over the power spectrum method when it comes to the calculation of the autocorrelation. Clearly in all cases it only takes the real space formalism a tiny fraction of a second to do the calculation. We however emphasize that this is only for the autocorrelation where the real space two dimensional integral \eqref{eq:orf_realspace} is easiest. As a matter of fact, elsewhere, the real space formalism is challenging to compute if not too often spoiled by numerical errors because of the poles of the integrand. We demonstrate this in Figure \ref{fig:orfTv1_PSvsRSF} for the ORF calculation in the simple case of near luminal tensor modes.} 

{This showcases one of the pragmatic advantages of the power spectrum method over the real space formalism. For an angular grid of eight points, the real space formalism took about three hundred seconds for this simple calculation with near luminal tensor modes, while the power spectrum method needed only a tenth of a second for the same task using only ten multipoles. Most impressively, for a grid of sixty points, the power spectrum calculation time took only a little bit longer than a tenth of second. The reason for this is that after the multipoles have been computed, the sum in \eqref{eq:orf1} can be readily performed for any angles, or alternatively the power spectrum allows control over the angular resolution with the minimum information, i.e., few multipoles are needed. On the other hand, the real space two dimensional integral \eqref{eq:orf_realspace} generally takes a long time to compute since it uses the same huge amount of information at small and large scales. This explains the computation times in Figure \ref{fig:orfTv1_PSvsRSF} for near luminal tensor modes.}

{The real space formalism performs worse in the more general case not only in terms of the computation times but also in the matter of numerical errors. Nonetheless these issues can be circumvented with enough computational power and that in fact the overlap reduction functions can be precomputed prior to stochastic gravitational wave background searches in pulsar timing array data analysis \cite{Cornish:2017oic, Chen:2021wdo}.}

{We strongly emphasize that the perks of the power spectrum method are not even limited to the control over the angular resolution, accuracy, and computation times. This algorithm unifies the stochastic gravitational wave background phenomenology in a pulsar timing array in a coherent few lines (Section \ref{sec:summary}) for all possible gravitational wave polarizations, putting together decades of work in pulsar timing array cosmology. We moreover expect that as more millisecond pulsars are utilized in pulsar timing arrays, thereby covering more sky and data points, the analysis would eventually resort to the power spectrum, as did cosmic microwave background science. In addition, the multipoles also give a direct route to the calculation of the higher moments of the correlation, or rather the variance of the overlap reduction function, as we have shown in \cite{Bernardo:2022xzl} not only for the Hellings-Downs curve but for astonishingly all possible modes. It is by these theoretical and practical standards that we see the power spectrum method a clear cut above the real space formalism for stochastic gravitational wave background phenomenology.}

We repeat the calculation for pulsars as far as $fD = 500$ or $D \sim 150$ parsecs, and can confirm similar observations. At this larger distance, however, we find that the power spectrum method becomes quite challenged by the extremely rapid oscillations in the integrand for large multipole numbers. The computation slows down considerably and the accuracy becomes less reliable, in addition to needing more multipoles, this time with $l \leq 3000$, for numerical convergence. This shows where the real space formalism, providing autocorrelation functions instantly, takes the advantage over the power spectrum method. The power spectrum even overestimates the autocorrelation function, which we associate to a breakdown of precision that the computation suffers with highly oscillating integrands. Nonetheless, the power spectrum method works well for nearby pulsar distances and that the real space formalism is always there to backup our calculations in this regime.

We make some final remarks on the drawbacks of the infinite distance limit, where the pulsars are at unreachable distances from the observer, and the power spectrum method. Table \ref{tab:GaafDinf} shows the autocorrelation calculated using the power spectrum and the real space formalism.

\begin{table}[h!]
    \centering
    \caption{Autocorrelation $\gamma_{aa}$ calculated using the power spectrum ($l \leq l_\text{max}$) and the real space formalism (RSF) in the infinite pulsar distance limit, $fD \rightarrow \infty$. The modes T, V, ST, and SL stand for `tensor', `vector', `scalar transverse', and `scalar longitudinal', respectively. {$t_{\rm eval}$ are the corresponding numerical evaluation times (in seconds) in a 12th Gen Intel Core i7-12700, 2100 Mhz Processor Computer.}}
    \begin{tabular}{|c|c|cc|cc|cc|} \hline
    mode & $\phantom{\dfrac{1}{1}}$ $v$ $\phantom{\dfrac{1}{1}}$ & $\phantom{gg}$ $\gamma_{aa}^{l \leq 30}$ $\phantom{gg}$ & $t_{\rm eval}$ (s) & $\phantom{gg}$ $\gamma_{aa}^{l \leq 1000}$ $\phantom{gg}$ & $t_{\rm eval}$ (s) & $\phantom{gg}$ $\gamma_{aa}^{\text{RSF}}$ $\phantom{gg}$ & $t_{\rm eval}$ (s) \\ \hline \hline
    \multirow{3}*{\phantom{ggg} T \phantom{ggg}} & $\phantom{\dfrac{1}{1}}$ $0.99$ $\phantom{\dfrac{1}{1}}$ & $1.08$ & $10^{-3}$ & $1.08$ & $2\times 10^{-3}$ & $2.17$ & 0.03 \\ \cline{2-8}
    & $\phantom{\dfrac{1}{1}}$ $0.50$ $\phantom{\dfrac{1}{1}}$ & $0.53$ & $10^{-3}$ & $0.53$ & $10^{-3}$ & $1.06$ & 0.01 \\ \cline{2-8}
    & $\phantom{\dfrac{1}{1}}$ $0.01$ $\phantom{\dfrac{1}{1}}$ & $0.47$ & $10^{-3}$ & $0.47$ & 0.03 & $0.97$ & $10^{-3}$ \\ \hline
    \multirow{3}*{\phantom{ggg} V \phantom{ggg}} & $\phantom{\dfrac{1}{1}}$ $0.99$ $\phantom{\dfrac{1}{1}}$ & $7.77$ & $10^{-3}$ & $7.77$ & 0.02 & $15.5$ & 0.03 \\ \cline{2-8}
        & $\phantom{\dfrac{1}{1}}$ $0.50$ $\phantom{\dfrac{1}{1}}$ & $0.67$ & $10^{-3}$ & $0.67$ & 0.02 & $1.34$ & 0.01 \\ \cline{2-8}
        & $\phantom{\dfrac{1}{1}}$ $0.01$ $\phantom{\dfrac{1}{1}}$ & $0.47$ & $10^{-3}$ & $0.47$ & 0.02 & $1.19$ & $10^{-3}$ \\ \hline
    \multirow{3}*{\phantom{ggg} ST \phantom{ggg}} & $\phantom{\dfrac{1}{1}}$ $0.99$ $\phantom{\dfrac{1}{1}}$ & $1.24$ & $10^{-3}$ & $1.24$ & 0.02 & $2.17$ & 0.03 \\ \cline{2-8}
    & $\phantom{\dfrac{1}{1}}$ $0.50$ $\phantom{\dfrac{1}{1}}$ & $15.4$ & $10^{-3}$ & $15.4$ & 0.02 & $1.06$ & 0.01 \\ \cline{2-8}
    & $\phantom{\dfrac{1}{1}}$ $0.01$ $\phantom{\dfrac{1}{1}}$ & $8.86 \times 10^7$ & $10^{-3}$ & $8.86 \times 10^7$ & 0.02 & $0.97$ & $10^{-3}$ \\ \hline
    \multirow{3}*{\phantom{ggg} SL \phantom{ggg}} & $\phantom{\dfrac{1}{1}}$ $0.99$ $\phantom{\dfrac{1}{1}}$ & $66.1$ & $10^{-3}$ & $66.2$ & 0.01 & $151$ & 0.03 \\ \cline{2-8}
    & $\phantom{\dfrac{1}{1}}$ $0.50$ $\phantom{\dfrac{1}{1}}$ & $31.0$ & $10^{-3}$ & $31.0$ & 0.01 & $1.26$ & 0.01 \\ \cline{2-8}
    & $\phantom{\dfrac{1}{1}}$ $0.01$ $\phantom{\dfrac{1}{1}}$ & $1.77 \times 10^8$ & $10^{-3}$ & $1.77 \times 10^8$ & 0.01 & $0.40$ & $10^{-3}$ \\ \hline
    \end{tabular}
    \label{tab:GaafDinf}
\end{table}

As expected, as the power spectrum profile merely drops sharply as the multipole number increases, the infinite distance power spectrum calculation misses about half of the total power that is contained in small angular scales, regardless of how many multipoles are included. This is reflected quite clearly for the tensor and vector modes in Table \ref{tab:GaafDinf} where the $\gamma_{aa}^{l \leq 30}$ and $\gamma_{aa}^{l \leq 1000}$ are the same at this reasonable level of precision. For the scalar polarizations, the situation worsens, now that the monopole and dipole spoils the power spectrum calculation through their dominance at low velocities. In this limit, in fact, it can be shown that the monopole and dipole behaves as $1/v^4$ and $1/v^2$, respectively, at low speeds $v \ll 1$ \cite{Qin:2020hfy}. Exactly, a straightforward integration for the monopole in the limit $fD \rightarrow \infty$ leads to $C_0^{\text{ST}}/(4\pi) = \sqrt{\pi}/\left(2 v^4\right) \sim 8.86 \times 10^7$ and $C_0^{\text{SL}}/(4\pi) = \sqrt{\pi}/v^4 \sim 1.77 \times 10^8$ as $v \sim 1/100$. This alone explains the unphysical numbers in Table \ref{tab:GaafDinf} for the scalar modes at half the speed of light and the nonrelativistic cases. A similar analytical calculation with the dipole can be done in this limit. Just the same, it appears that only the real space formalism is trustworthy for the autocorrelation in the infinite distance limit. We take these considerations to add more case to factoring in finite pulsar distances in pulsar timing array analysis. {Nonetheless we can utilize analytical expressions for the integrals in the infinite distance limit, as we derived in the previous sections, to aid in the numerical evaluation and make the computation times competitive with the real space formalism counterparts.}

In a future work, we look forward to a merging of the power spectrum method and the real space formalism for an accurate calculation of the overlap reduction function in a pulsar timing array, in the same way post Newtonian gravity and numerical relativity are utilized in studying gravitational waves from compact binaries. This will become relevant as pulsar pairs of subdegree separation are observed, something which we anticipate in future pulsar timing array missions.

\section{Conclusions}
\label{sec:conclusions}

We have derived the overlap reduction functions of an isotropic stochastic gravitational wave background sourced by tensor, vector, and scalar metric polarizations, for finite pulsar distances and subluminal velocities. This reveals for one the importance of keeping the pulsars at realistic finite distances, particularly, in letting all modes be well defined and keeping the power in small scales, and in future pulsar timing array data where millisecond pulsar pairs may be of subdegree separations. Our technical results prepare the general data analysis of the various possible metric polarizations which may be anchored in the nanohertz gravitational wave sky, such as we have demonstrated in \cite{Bernardo:2022vlj}. However the results may be, this would be complementary to the picture provided by ground and space based gravitational wave observatories about our Universe and the matter that lives, and lived, within.

The power spectrum calculation is an efficient way of obtaining the pulsar timing array observables, in addition to providing an independent check of the real space formalism which is known to be technically challenged by poles in numerical integration over an angular domain. Even so, the power spectrum calculation and the real space formalism for obtaining the overlap reduction function can be complimentary, particularly in future data sets with pulsar pairs of subdegree separation, which will challenge the power spectrum calculation but not quite the real space formalism. We have shown this for the autocorrelation function, describing pulsar pairs along the same line of sight or the correlation of a pulsar with itself, where at least a few thousand multipoles were needed to ensure the numerical convergence of the sum in the power spectrum calculation but which was a quick calculation using the real space formalism. A merger of these two methods, much in the same way gravitational waveform analyses rely on post Newtonian and numerical relativity calculations, would be something to look forward to in the pulsar timing community.

This work sets up several future directions. First, it would be interesting to see whether the tensor modes by themselves, if they are freed from the light cone, could match statistically the present pulsar timing array data. If so, this hints at dispersive gravitational waves that only so happen to be luminal in the ground based detectors' frequency band of about a hundred hertz. Second, following on \cite{Qin:2020hfy, Bernardo:2022vlj}, in theories, the various metric polarizations are expected to be constrained by the theory parameters rather than just independently contributing to the overall signal. This calls on alternative gravity theorists to setup these theoretical constraints which would also reduce the parameter space and lift potential degeneracy of the models for data analysis. Lastly, building on \cite{Liu:2022skj}, it remains to setup the general formalism for scalar and vector polarizations, as well as tensors off the light cone, for studying the anisotropies in the stochastic gravitational wave background. It may take a while for pulsar timing science to mature to this level of sensitivity, but this takes to a different level, hinting at natural clocks and nonGaussianities in the primordial universe \cite{Pol:2022sjn, Bodas:2022zca, Dimastrogiovanni:2022afr}.

\acknowledgments

This work was supported in part by the Ministry of Science and Technology (MOST) of Taiwan, Republic of China, under Grant No. MOST 110-2112-M-001-036.

\appendix

\section{Gravitational wave polarization basis}
\label{sec:gw_pol_tensors}

For a gravitational wave propagating along the $\hat{\Omega}$ direction, the polarization basis tensors can be expressed as \cite{Boitier:2021rmb}
\begin{eqnarray}
\varepsilon^{+} &=& \hat{m} \otimes \hat{m} - \hat{n} \otimes \hat{n} \,, \\
\varepsilon^{\times} &=& \hat{m} \otimes \hat{n} + \hat{n} \otimes \hat{m} \,, \\
\varepsilon^{x} &=& \hat{m} \otimes \hat{\Omega} + \hat{\Omega} \otimes \hat{m} \,, \\
\varepsilon^{y} &=& \hat{n} \otimes \hat{\Omega} + \hat{\Omega} \otimes \hat{n} \,, \\
\varepsilon^{\text{ST}} &=& \hat{m} \otimes \hat{m} + \hat{n} \otimes \hat{n} \,, \\
\varepsilon^{\text{SL}} &=& \sqrt{2} \hat{\Omega} \otimes \hat{\Omega} \,,
\end{eqnarray}
where $\left( \hat{m}, \hat{n}, \hat{\Omega} \right)$ for an orthonormal basis. The $\left(\varepsilon^{+}, \varepsilon^{\times} \right)$ stand for the transverse-traceless tensor modes, $\left(\varepsilon^{x}, \varepsilon^{y} \right)$ for the vector modes, and $\left(\varepsilon^{\text{ST}}, \varepsilon^{\text{SL}} \right)$ for the scalar modes.

In practice, it is useful to orient $\hat{\Omega}$ along the $\hat{z}$-direction. In this case, the orthonormal basis $\left( \hat{m}, \hat{n}, \hat{\Omega} \right)$ may be written as
\begin{eqnarray}
\hat{m} &=& \cos \varphi \ \hat{x} + \sin \varphi \ \hat{y} \,, \\
\hat{n} &=& -\sin \varphi \ \hat{x} + \cos \varphi \ \hat{y} \,, \\
\hat{\Omega} &=& \hat{z} \,.
\end{eqnarray}
With the Cartesian basis $\left( \hat{x}, \hat{y}, \hat{z} \right)$, the polarization tensors can be identified to be
\begin{equation}
    \varepsilon^{+} = 
    \left(
    \begin{array}{ccc}
    \cos(2\varphi) & \sin(2\varphi) & 0 \\
    \sin(2\varphi) & -\cos(2\varphi) & 0 \\
    0 & 0 & 0 
    \end{array}
    \right) \,,
\end{equation}
\begin{equation}
    \varepsilon^{\times} = 
    \left(
    \begin{array}{ccc}
    -\sin(2\varphi) & \cos(2\varphi) & 0 \\
    \cos(2\varphi) & -\sin(2\varphi) & 0 \\
    0 & 0 & 0 
    \end{array}
    \right) \,,
\end{equation}
\begin{equation}
    \varepsilon^{x} = 
    \left(
    \begin{array}{ccc}
    0 & 0 & \cos \varphi \\
    0 & 0 & \sin \varphi \\
    \cos \varphi & \sin \varphi & 0 
    \end{array}
    \right) \,,
\end{equation}
\begin{equation}
    \varepsilon^{y} = 
    \left(
    \begin{array}{ccc}
    0 & 0 & -\sin \varphi \\
    0 & 0 & \cos \varphi \\
    -\sin \varphi & \cos \varphi & 0 
    \end{array}
    \right) \,,
\end{equation}
\begin{equation}
    \varepsilon^{\text{ST}} = 
    \left(
    \begin{array}{ccc}
    1 & 0 & 0 \\
    0 & 1 & 0 \\
    0 & 0 & 0 
    \end{array}
    \right) \,,
\end{equation}
and
\begin{equation}
    \varepsilon^{\text{SL}} = 
    \sqrt{2} \left(
    \begin{array}{ccc}
    0 & 0 & 0 \\
    0 & 0 & 0 \\
    0 & 0 & 1 
    \end{array}
    \right) \,.
\end{equation}

\section{Triple spherical harmonics integral and the Wigner-3j symbol}
\label{sec:3Y3j}

We put down identities on spin weighted spherical harmonics $\,_s Y_{lm}\left(\hat{n}\right)$ \cite{Weinberg, Arfken}.

First, for $s = 0$, $\,_s Y_{lm}\left(\hat{n}\right)$ reduces to the spherical harmonic $ Y_{lm}\left(\hat{n}\right)$. In general, $\,_s Y_{lm}\left(\hat{n}\right)$ satisfies the orthogonality relation
\begin{equation}
    \int_{S^2} d\hat{n} \,_s Y_{lm}^*\left(\hat{n}\right) \,_s Y_{l'm'}\left(\hat{n}\right) = \delta_{ll'} \delta_{mm'} \,,
\end{equation}
and completeness relation
\begin{equation}
    \sum_{lm} \,_s Y^*_{lm}\left(\hat{n}\right) \,_s Y_{lm}\left(\hat{n}' \right) = \delta^{(2)}\left( \hat{n} - \hat{n}' \right) = \delta\left(\phi - \phi'\right) \delta\left(\cos \theta - \cos \theta' \right) \,.
\end{equation}
This also satisfies the conjugate identity
\begin{equation}
    \,_s Y_{lm}^*\left( \hat{n} \right) = (-1)^{s + m} \,_{-s}Y_{l-m}\left( \hat{n} \right) \,.
\end{equation}

We progress in the text with triple spherical harmonics identity
\begin{equation}
    \int d \hat{e} \,_{s_1} Y_{l_1 m_1}\left(\hat{e}\right) \,_{s_2}Y_{l_2 m_2}\left(\hat{e}\right) \,_{s_3}Y_{l_3 m_3}\left(\hat{e}\right) = \sqrt{\dfrac{(2l_1+1)(2l_2+1)(2l_3+1)}{4\pi}} \left( \begin{array}{ccc}
    l_1 & l_2 & l_3 \\
    -s_1 & -s_2 & -s_3
    \end{array} \right)
    \left( \begin{array}{ccc}
    l_1 & l_2 & l_3 \\
    m_1 & m_2 & m_3
    \end{array} \right) \,,
\end{equation}
where $\left(\begin{array}{ccc} a & b & c \\ d & e & f \end{array}\right)$ is the Wigner-3j symbol. The 3j symbol vanishes unless $|l_1 - l_2| < l_3 < l_1 + l_2$ and $m_1 + m_2 + m_3 = 0$. Further, if $m_1 = m_2 = m_3 = 0$, then $l_1 + l_2 + l_3$ must be an even integer. It also satisfies a reflection property
\begin{equation}
    \left( \begin{array}{ccc}
    l_1 & l_2 & l_3 \\
    m_1 & m_2 & m_3
    \end{array} \right)
    = (-1)^{l_1 + l_2 + l_3}
    \left( \begin{array}{ccc}
    l_1 & l_2 & l_3 \\
    -m_1 & -m_2 & -m_3
    \end{array} \right) \,.
\end{equation}

\section{Another real space formalism}
\label{sec:another}

Provided a passing gravitational wave in the direction $\hat{k}$, the redshift fluctuation is also often considered as the Shapiro time delay \cite{NANOGrav:2021ini}:
\begin{equation}
\label{eq:redshift_realspace}
z(t) = \dfrac{\hat{e}^i \otimes \hat{e}^j}{2\left( 1 + v \hat{k} \cdot \hat{e} \right)} \left( h_{ij}^e - h_{ij}^p \right) \,,
\end{equation}
where $h_{ij}^e = h_{ij}\left(t, \vec{0}\right)$, the earth term, is the metric perturbation evaluated on earth where the pulse is received, and $h_{ij}^p = h_{ij}\left( t - D , D \hat{e} \right)$, the pulsar term, is evaluated at the pulsar during emission. Substituting \eqref{eq:gw_general}, the redshift fluctuation for a pulsar $a$ can be simplified as
\begin{equation}
\label{eq:redshift_altreal}
    z_a(t) = \sum_A \int_{-\infty}^\infty df \int_{S^2} d\hat{k} \ \tilde{h}_A\left(f, \hat{k}\right) F_a^A \left( \hat{k} \right) e^{-2\pi i f t} U_a\left( f, \hat{k} \right) \,,
\end{equation}
where $F_a$ are the antenna pattern functions \eqref{eq:antenna_functions} and $U_a$ is given by \eqref{eq:Uadef}. Substituting \eqref{eq:redshift_altreal} into \eqref{eq:timing_residual}, one gets to the two point pulsar timing residual correlation function \eqref{eq:twopoint_realspace} and the overlap reduction function \eqref{eq:orf_realspace}.

\section{A brief review of the overlap reduction function}
\label{sec:orfs_review}

We briefly review the well established results about the overlap reduction function (see e.g. \cite{NANOGrav:2021ini}).

We start with the standard one, that is, due to the transverse traceless tensor polarizations predicted by general relativity. The overlap reduction function is given by
\begin{equation}
    \Gamma_{ab}^{\text{TT}} = \Gamma_{ab}^+ + \Gamma_{ab}^\times \approxeq \dfrac{\delta_{ab}}{2} + C\left(\zeta_{ab}\right) \,,
\end{equation}
where $\zeta_{ab}$ is the angular separation of two pulsars, and $C\left(\zeta_{ab}\right)$ is the Hellings-Downs curve \cite{Hellings:1983fr}:
\begin{equation}
    C\left(\zeta_{ab}\right) = \dfrac{3}{2} \left( \dfrac{1}{3} + \left( \dfrac{1 - \cos \zeta_{ab}}{2} \right) \left[ \ln \left( \dfrac{1 - \cos \zeta_{ab}}{2} \right) - \dfrac{1}{6} \right] \right) \,.
\end{equation}
Provided tensor modes propagating at the speed of light, we expect the stochastic gravitational wave background signal to be given by the Hellings-Downs correlation.

There are also phenomenological ones, that are not necessarily due to gravitational degrees of freedom, but nonetheless present a competitive signal to noise ratio in the current data \cite{NANOGrav:2021ini}. These are the gravitational wave like monopole and dipole:
\begin{equation}
    \Gamma_{ab}^{\text{GW mon}} = \dfrac{\delta_{ab}}{2} + \dfrac{1}{2} \,,
\end{equation}
and
\begin{equation}
    \Gamma_{ab}^{\text{GW dip}} = \dfrac{\delta_{ab}}{2} + \dfrac{\cos \zeta_{ab}}{2} \,.
\end{equation}
The gravitational wave monopole particularly has a significant signal to noise ratio, compared with the gravitational wave dipole and the Hellings-Downs correlation, in the present data set.

We move on to non-Einsteinian polarization modes on the light cone, and considering infinite pulsar distances. The scalar transverse polarization (also often referred to as the `breathing' mode) leads to the overlap reduction function \cite{Chamberlin:2011ev}
\begin{equation}
    \Gamma_{ab}^{\text{ST}} \approx \dfrac{\delta_{ab}}{2} + \dfrac{1}{8} \left( 3 + \cos \zeta_{ab} \right) \,.
\end{equation}
On the other hand, for the scalar longitudinal modes, the overlap reduction function cannot be evaluated analytically for arbitrary pulsar pair angular separations. But the more prominent issue is that this cannot be defined for infinite pulsar distances. Keeping the pulsars at a finite distance $fD \gg 1$, from the observer, the autocorrelation function can be shown to be linearly divergent \cite{Chamberlin:2011ev},
\begin{equation}
     \Gamma_{aa}^\text{SL} \sim \dfrac{3\pi^2}{4} fD - 3 \ln \left( 4\pi fD \right) + \dfrac{37}{8} - 3 \gamma_\text{E} \,,
\end{equation}
where $\gamma_\text{E}$ is Euler's constant. A similar situation arises for the vector modes, whereas the overlap reduction function can be determined to be \cite{2008ApJ...685.1304L}
\begin{equation}
    \Gamma_{ab}^\text{V} = \Gamma_{ab}^{(\text{V})_x} + \Gamma_{ab}^{(\text{V})_y} \approx 3 \log \left( \dfrac{2}{1 - \cos \zeta_{ab}} \right) - 4 \cos \zeta_{ab} - 3 \,,
\end{equation}
the autocorrelation function \cite{2008ApJ...685.1304L}
\begin{equation}
    \Gamma_{aa}^\text{V} \sim 6 \ln \left( 4 \pi fD \right) - 14 + 6 \gamma_\text{E} \,
\end{equation}
becomes undefined, diverges logarithmically, in the infinite distance limit, albeit not as strongly as the scalar longitudinal polarization.


\bigskip


%

\end{document}